\documentclass[a4paper,12pt]{article}

\usepackage{xcolor}
\usepackage{newtxtext,newtxmath}
\usepackage{graphicx}
\usepackage{caption}
\usepackage{float}
\usepackage{multirow} 
\usepackage{makecell}

\usepackage[]{hyperref}
\hypersetup{
	colorlinks=true,
	linkcolor=blue,
	filecolor=violet,     
	urlcolor=blue,
	citecolor=blue
}
\usepackage{epsfig}%

\newcommand{\nb}{\ensuremath{ \nabla }}

\newcommand{\m}{\ensuremath{{\mu \nu}}}

\newcommand{\p}{\ensuremath{\partial{}}}

\newcommand{\pp}{\ensuremath{\mathcal{P}_{i}}}

\title{\bf Dynamical stability in models where dark matter and dark energy are non-minimally coupled to  curvature}
\bigskip

\author{Saddam Hussain $^{\$}$, Anirban Chatterjee $^\star$, Kaushik Bhattacharya$^\dagger$  
\noindent\thanks{$^\$$msaddam@iitk.ac.in, $^\star$anirbanc@iitk.ac.in, $^\dagger$ kaushikb@iitk.ac.in}\\
\normalsize
Department of Physics, Indian Institute of Technology, Kanpur\\ 
\normalsize
Uttar Pradesh 208016, India
}

\begin{document}
\maketitle
\begin{abstract}
This work explores the dynamical stability of cosmological models where dark matter and dark energy can non-minimally couple to spacetime (scalar) curvature. Two different scenarios are presented here. In the initial case, only dark matter sector is coupled to curvature in the presence of a quintessence scalar field. In the second case both dark matter and the quintessence field are coupled to curvature. It is shown that one can get an accelerating expansion phase of the universe in both the cases. The nature of the fixed points shows that there can be stable or unstable phases where the curvature coupling vanishes and dark energy and dark matter evolve independently. On the other hand there can be stable accelerating expansion phases where both the components are coupled to curvature.   
\end{abstract}

	
\section{Introduction}

Consistent evidence for the current pace of cosmic expansion has been accumulating from various observational results \cite{1998fmf, 1998vns, SDSS, Planck}  since 1998. Dark energy (DE) is an exotic fluid that has been suggested to be responsible for this phenomenon, its negative pressure may adequately account for the observed expansion rate of the universe. This fluid is estimated to make up around 70\% of the entire energy content of the Universe \cite{Copeland:2006wr,Clifton:2011jh,Bahamonde:2017ize}. Currently, the most widely accepted cosmological model is the $\Lambda$-CDM model \cite{Peebles:2002gy}, which consists of a cosmological constant dark energy (DE) source and a dark matter (DM) component necessary to enable the development of the structure conceivable in the Universe. There are some theoretical issues with this model, which has led some to propose the idea of dark energy produced by a scalar field \( \phi \), within the framework of general relativity \cite{Wetterich:1987fm,Peebles}. Despite their seeming simplicity, models based on scalar fields may generate intricate and detailed phenomenologies while still generating predictions that can be tested against existing data \cite{Planck}. Most simply, DE is produced by  a canonical scalar field, the quintessence field, that has no interactions with anything else in the Universe \cite{Tsujikawa:2013fta,Carroll:1998zi}. Later on it was seen that there is no essential reason to make this assumption, and in the simplest extension, the scalar field \(\phi\) is permitted to interact with the matter sector \cite{Ellis:1989as,Amendola:1999qq,Amendola:1999er,Nunes:2000ka,Gumjudpai:2005ry,Boehmer:2008av,Barros:2018efl}. In these cases, the coupling has been introduced at the level of the continuity equation by hand as $ \nb_{\mu} T^{\m}_{\rm DM} = -Q^{\nu}, \ \nb_{\mu} T^{\m}_{\phi} = Q^{\nu} $, where \( Q^{\nu} \) is the interaction term. In these cases, the total energy-momentum tensor is conserved; however, the field and fluid exchange  energy via \( Q^{\nu}\). One of the fundamental reasons for studying this kind of extended coupled scalar field model is to address the shortcomings of the $\Lambda$-CDM model; precisely, the Hubble tension and the amplitude $S_8$ of matter density \cite{DiValentino:2021izs,Krishnan:2020obg,Davis:2019wet,Abdalla:2022yfr,Krishnan:2020vaf,Aloni:2021eaq} difference between high and low-redshift data \cite{Planck:2018nkj,Planck:2015fie,Riess:2016jrr,Hildebrandt:2016iqg}. Although this approach seems to fit with cosmological data and reduces the Hubble and \( S_8\) tensions \cite{DiValentino:2019ffd} but in Ref. \cite{Tamanini:2015iia}, it has been argued that the covariant approach of introducing the interaction term leads to several theoretical problems. These issues have direct consequences in the study of cosmological perturbations. Such hints dictate the study of the interaction of DM-DE from the Lagrangian approach, which yields the covariant energy-momentum tensors and gives a consistent framework to study cosmological perturbations.

Several studies have been carried out \cite{Pourtsidou:2013nha,Boehmer:2015kta,Boehmer:2015sha,Chatterjee:2021ijw,Bhattacharya:2022wzu,Koivisto:2015qua,Kase:2019veo} in which the authors have studied how the fluid component interacts with a scalar field sector. In these models the fluid generally represents the DM sector and the field part represents the DE sector. The action of a relativistic fluid was first introduced in \cite{Schutz:1977df} and further developed by Brown \cite{Brown:1992kc}. The action contains fluid energy density, particle flux number, and Lagrangian multipliers. When non-minimal field-fluid interaction is incorporated, the two dark sectors directly interact while individually both the components remain minimally coupled to gravity. Nevertheless, this coupling can be extended, as argued by Bettoni et al. in \cite{Bettoni:2015wla}, where the DM sector interacts with the curvature producing a non-minimally coupled (NMC) system. The idea of non-minimal coupling of matter with curvature \cite{Brans:1961sx,Sen:2000zk,Bertolami:2008ab,Bertolami:2011rb,Bertolami:2007gv,Harko:2014pqa,Gomes:2019tmi,Bettoni:2011fs} is very rich. In previous cases, coupling with gravity have introduced modification in the gravitational or fluid sectors. The other approach discussed in \cite{Bettoni:2015wla}, generalizes in such a way that any modifications in the matter sector can induce a significant change in the gravitational sector. The curvature couplings introduced are of two types: in one case the fluid variable is directly coupled to  the scalar curvature, $R$, in the other case the fluid variable is coupled to  or $R_{\m} U^{\mu} U^{\nu}$.  These couplings are called conformal and disformal couplings. Here \( R\) is a Ricci scalar, $ R_{\m}  $ is a Ricci tensor and \( U^{\mu}\) is the four velocities of the fluid. 
	
In this paper we study the cosmological dynamics of models which involve conformal coupling of the fluid in addition to the Quintessence field using the dynamical system approach. The main motivation for the present work is related to the question: How does the DE sector affect cosmological models where DM is conformally coupled to gravity? Can we find stable accelerating expansion phases of the late universe in presence of conformal coupling? Does the stable accelerated expansion phase show nonzero gravitational coupling? We will answer these questions in this work. It is seen that depending upon the conformal coupling one may have qualitatively different classes of stable accelerated expansion. Some of these stable expansion phases are conformally decoupled and essentially represent a  two component universe: the two components being DE and DM. For a different conformal coupling we can also have stable accelerated expansion where DM is always coupled to gravity. The latter kind of models are cosmologically interesting as in these cases the theory of structure formation becomes more involved. In this work, we have also formulated a more adventurous model where all the matter components, which includes the DM and DE sector, together couple non-minimally to gravity via a conformal like coupling. Doing so we have generalized the model previously used in Ref.~\cite{Bettoni:2015wla}. Our work primarily uses the dynamical system methods to find out the stable accelerated expansion phases of the late universe. The dynamical system approach is one of the crucial techniques used to determine the stability and global dynamical evolution of the system. The autonomous system of equations are constructed by choosing a set of dimensionless variables. The critical points in various models are obtained, and a Jacobian matrix related to the autonomous systems are constructed by linearizing the autonomous equations around the fixed points. The nonzero real part of the eigenvalues of this matrix determines the stability of the system. However, this technique fails if any eigenvalue is zero. In that case, more rigorous analytical techniques such as center manifold theorem or Lyapunov stability have to be applied \cite{Rendall:2001it,Boehmer:2011tp,Dutta:2017wfd,Coley:2003mj,blackmore2011nonlinear}.
	
The structure of the paper is as follows. In section \ref{sec:cuv_fluid}, we present the basic analysis of curvature coupling.  We present models involving conformal coupling between DM sector and curvature in presence of an additional minimally coupled canonical scalar field in section \ref{sec:quit_minimally}. In section \ref{sec:quint_nonminimally} we generalize our previous result and formulate a theory where both the DM and DE are simultaneously coupled to scalar curvature. We conclude or work with some relevant discussion on the results in section \ref{sec_conclusion}.

\section{Dynamics of Non-minimally coupled Curvature-Fluid}
\label{sec:cuv_fluid}

The action of a non-minimally coupled fluid with the curvature as introduced in Ref.\cite{Bettoni:2015wla} is
\begin{equation}\label{betoni_action}
S_{\rm cf} = \int_{\Omega} d^4 x \left[\sqrt{-g}\dfrac{R}{2 \kappa^2} -\sqrt{-g} \, \rho(n,s) +
J^\mu(\varphi_{,\mu} + s\theta_{,\mu} + \beta_A\alpha^A_{,\mu}) \\
+ \sqrt{-g}\,\alpha f(n,s) \frac{R}{2\kappa^2}\right]\,.
\end{equation}
Here the first term is Einstein-Hilbert action, where \(g\) is the determinant of the metric \(g_{\m}\), $\kappa^2 = 8 \pi G$ and \(R\) is a Ricci scalar. The second term quantifies action corresponding to the relativistic fluid in which \( \rho\) is energy density of the fluid,  \(n\) is particle number density, \( s\) entropy density per particle, and \( ( \varphi, \ \theta, \ \beta_A, \  \alpha^{A} )\) are the Lagrangian multipliers. The commas $ \varphi_{, \mu} \equiv \p_{\mu} \varphi $, are the partial derivative with respect to space-time coordinates. The last term signifies the interaction which  couples the fluid  with the curvature with a dimensionless coupling constant $ \alpha $. The particle flux number is:
\begin{equation}\label{}
J^{\mu} = \sqrt{-g } n U^{\mu }, \ |J| = \sqrt{-g_{\m} J^{\mu} J^{\nu}}, \ n = \frac{|J|}{\sqrt{-g}}, \ U^{\mu} = (1, \vec{0})\,. 
\end{equation}
Here the fluid 4-velocity \( U^{ \mu}\) satisfying $U^{\mu} U_{\mu} = -1$. The variation corresponding to the fluid variables is shown in appendix \ref{appen:1}. Varying the action with respect to \( g^{\m}\) gives the modified Einstein field equation as:
\begin{multline}\label{curvature fluid field equaion_1}
\dfrac{1}{\kappa^2}\left[R_{\m} - \dfrac{1}{2} R g_{\m} + \alpha f R_{\m} + \alpha (g_{\m} \nb_{\sigma} \nb^{\sigma} f - \nb_{\mu} \nb_{\nu} f) \right]  = \left( \rho -  \dfrac{1}{2 \kappa^2} \alpha f R \right) U_{\mu} U_{\nu}   \\
+ \left[n \rho_{,n} - \rho - \dfrac{\alpha R}{2 \kappa^2} (n f_{,n} -f)\right]  (U_{\mu} U_{\nu} + g_{\m} )\,.
\end{multline}
We have written the above equation in a standard form as $ \frac{1}{\kappa^2} G_{\m} =  T_{\m }^{f} + T_{\m}^{\rm int}  $, where the stress tensor is defined as $ T_{\m} = -\frac{2}{\sqrt{-g}} \frac{\delta S}{\delta g^{\m}} $. Comparing with the stress tensor of the perfect fluid $ T_{\m}^f = \rho U_{\mu} U_{\nu} + P (U_{\mu} U_{\nu} + g_{\m} ) $, we can write the effective energy density and pressure as \(\rho_{\rm tot}, P_{\rm tot}\) respectively. For homogeneous and isotropic background line element can be written as:
\begin{equation}\label{}
ds^2 = -dt^2 + a(t) ^2 d\vec{x}^2\,.
\end{equation}
Here the Friedmann equations can be expressed as: 
\begin{eqnarray}
3 H^2 &=& \dfrac{\kappa^2 \rho }{1 - 3 \alpha n f_{,n} + \alpha f} = \dfrac{\kappa^2 \rho }{1 - 3 \alpha \mathcal{P}_{i} - 2 \alpha f },\label{eq:first_fiedmann}\\
2 \dot{H} &=& \dfrac{\kappa^2 P_{M}}{ -1 + 2 \alpha f + 3 \alpha \mathcal{P}_{i}} - \dfrac{\kappa^2 \rho (1 - \alpha \mathcal{P}_{i} + 3 \alpha c_{i}^2 (\mathcal{P}_{i} + f))}{(-1 + 2 \alpha f + 3 \alpha \mathcal{P}_{i})^2}\,. \label{eq:frd_2}
\end{eqnarray}
The interaction pressure \( P_{\rm int}\) and the variable $c_i$
are defined as:
\begin{equation}\label{int_pressure}
  P_{\rm  int} = \frac{\alpha  R}{2 \kappa^2} \mathcal{P}_{i}, \quad	\mathcal{P}_{i} = n f_{,n} - f\,,\quad  c_{i}^2 f_{,n} = n f_{,nn}\,,
\end{equation}
where $\mathcal{P}_{i}$ is the dimensionless variable proportional to the interaction pressure. Rewriting Eq.~\eqref{curvature fluid field equaion_1} as $ G_{\m} = \kappa^2(T_{\m}^{1} + T_{\m}^{2}) $, where $ G_{\m} = R_{\mu \nu} - 1/2 R g_{\m} $, we have
\begin{equation}\label{reT1}
T_{\m}^1 = \left( \rho -  \dfrac{1}{2 \kappa^2} \alpha f R \right) U_{\mu} U_{\nu}   
+ \left[ n \rho_{,n} - \rho - \dfrac{\alpha R}{2 \kappa^2} (n f_{,n} -f)   \right]  (U_{\mu} U_{\nu} + g_{\m} )\,,
\end{equation}
and 
\begin{equation}\label{reT2}
T_{\m}^2 =\dfrac{-1}{ \kappa^2}	\bigg[   \alpha f R_{\m} + \alpha (g_{\m} \nb_{\sigma} \nb^{\sigma} f - \nb_{\mu} \nb_{\nu} f)\bigg]\,. 
\end{equation}
Taking into account the Bianchi identity, the covariant derivative of the Einstein tensor vanishes, $\nb^{\mu} G_{\m} = \kappa^2 \nb^{\mu} \left[ T_{\m}^{1} + T_{\m}^{2}\right] = 0$. In a FLRW metric, the covariant derivative of the redefined stress tensor will produce: 
\begin{eqnarray}
\nb_{\mu} T^{\m}_1 &=& \dfrac{-\alpha}{2\kappa^2} f \nb_{0}R\,,\nonumber\\
\nb_{\mu} T^{\m}_2 &=& 	\dfrac{-\alpha}{\kappa^2} \left[ 9 H (nf_{,n} -f) \dfrac{\ddot{a}}{a} - 3f \left( \dfrac{\dddot{a}}{a} - \dfrac{\ddot{a}}{a^2}\dot{a}\right)+ 3 H f \left(2 H^2 + \frac{\ddot{a}}{a}\right) - 9 \p_{0} (H^2 f_{,n } n) \right]\,.
\nonumber  
\end{eqnarray}
To derive the following, we have used these constrained \( \dot{n} + 3 nH =0\), \( \dot{s} = 0 \), $ \nb_{\mu } U^{\mu} = 3H $, $ U_{\lambda}U^{\lambda} = -1 $ and $ U^{\mu} U_{\lambda} \nb_{\mu}U^{\lambda} = 0 $. We impose the conservation condition: 
\begin{equation}\label{}
\nb_{\mu} (T^{\m}_1 + T^{\m}_{2}) = \dfrac{-6 \alpha H }{2 \kappa^2} \left[ -3 n f_{,n} \dfrac{\ddot{a}}{a}+ 9 n^2 H^2 f_{,nn} + 15 n H^2 f_{,n} \right] =0\,.
\end{equation}
This yields the interaction parameter, $c_i$, as:
\begin{equation}\label{fluid equation}
c_{i}^2  = \dfrac{ \dot{H} - 4 H^2}{3 H^2}\,.
\end{equation}
The additional constrain \( \dot{n} + 3 nH =0\), \( \dot{s} = 0 \), allow us to write the conserved quantity for the fluid and interaction along the flow line, 
\begin{equation}\label{conserved_interaction}
\begin{split}
\dot{\rho}+3H (\rho + P)=0\,,\quad \dot{f} + 3 H(f + \mathcal{P}_{i}) = 0\,.
\end{split}
\end{equation}
Hence using Eq.~\eqref{fluid equation}, the second Friedmann equation can be rewritten as: 
\begin{equation}\label{2nd_fried_modified}
\dot{H} = \dfrac{\kappa^2 P_M}{(-2 + \alpha f + 3 \alpha \mathcal{P}_{i} )} - \dfrac{\kappa^2 \rho\left[1- \alpha \mathcal{P}_{i} - 4 \alpha (\mathcal{P}_{i} + f)\right]}{(-1 + 2 \alpha f + 3 \alpha \mathcal{P}_{i}) (-2 + \alpha f + 3 \alpha \mathcal{P}_{i})}\,.
\end{equation}
We will use these basic equations in the phenomenological models we study in this paper. 


{

\subsection{Dynamical analysis of curvature-fluid system \label{fluid_curv_dynamics}}	

In this subsection we will explore the dynamics, discussed before, using the following form of the interactions:
 \begin{equation}
 	 f(n,s) =  (I) \  M^{-4 \beta }\rho^{\beta}(n,s),\quad (II) \  \exp(M^{-4 \beta }\rho^{\beta}(n,s))
 \end{equation}
 The interaction consists of the fluid density $\rho$, a mass-dimensional constant $M$, and a dimensionless parameter $\beta$. These interaction models are chosen because they are perhaps the simplest and workable models one can use in the present case. To analyze the dynamics of the system, we will select first dimensionless variables as:
   	\begin{equation}\label{eq:dyn_curv_fluid}
  	\quad z=f\,, \quad \sigma^2 = \dfrac{\kappa^2 \rho}{3 H^2}\,.
  \end{equation} 
 With the choice of the dynamical variable \(z\) and first Friedmann equation Eq.~\eqref{eq:first_fiedmann}, the dynamics of the system can be analyzed only by the autonomous equation in \(z\). The effective equation of state can be written as 
 \begin{equation}\label{}
 -\dfrac{2 \dot{H}}{3 H^2}  =	\omega_{\rm tot} = -1 - 2 \left[ \frac{\omega \sigma^2}{(-2+ 3 \alpha \mathcal{P}_{i} +  \alpha z)} - \frac{\sigma^2  (1-\alpha \mathcal{P}_{i} - 4\alpha (\mathcal{P}_{i} + z))}{(-1+ 3 \alpha \mathcal{P}_{i} + 2 \alpha z)(-2+ 3 \alpha \mathcal{P}_{i} +  \alpha z)}\right]\,.
 \end{equation}

 \begin{table}[t]
 
 		\centering
 		\begin{tabular}{cccccc}
 			\hline
 			Models & $ \sigma^2 $ & $\pp$ & \(z\) & $\sigma^2 $ & $\omega_{\rm tot}$ \\
 			\hline
 			\hline
 			$ \rm{I} $ & $ 1-\alpha z (2+3(\beta(1+\omega)-1)) $& $ z \left(\beta(\omega + 1) - 1\right) $ & $ 0 $& $1$ & 0  \\
 			\hline
 			
 			$ \rm{II} $ & $ 1- 3 \alpha \beta z (1+\omega) \ln|z| + \alpha z $& $ z\left[\beta (\omega + 1) \ln|z| -1\right] $ & $(0,1)$ & $(1, 1+ \alpha) $ & $(0,0)$ \\
 			\hline
 		\end{tabular}
 		\caption{  The critical points and their nature for the coupled curvature-fluid system for two different interaction models.  }
 		\label{tab:model_cric_example}
 	
 \end{table}
 
The autonomous equation for Model I is 
 \begin{equation}
 	z' \equiv \frac{dz}{H dt} = -3 \beta z (1+\omega)\label{z_prime_cf1}
 \end{equation}
and for Model II:  
 \begin{equation}
 	z' = -3 \beta z (1+\omega) \ln|z|\, . \label{z_prime_cf2}
 \end{equation}
 \noindent where prime stands for differentiation with respect to $H\ dt  \equiv dN$. These critical points \(( z)\) for the two models are tabulated in Tab.[\ref{tab:model_cric_example}]. In addition to the  critical points, we have also evaluated the fluid fraction density \((\sigma^2)\) and pressure parameter $\pp$ corresponding to these models at the critical points. Model I generates negligible interaction at the critical point for any $\beta$ and the fluid density dominates near the fixed point.  At the fixed point the equation of state (EoS) is zero, denoting a matter-dominated phase. This demonstrates that the interaction becomes negligible when matter density dominates. On the other hand, Model II produces two critical points \((z \equiv (0,1))\), where the system at point at \(z = 0\) has similar properties as it had in Model I. At the fixed point \( z \equiv 1\) the fluid density depends on the model parameter \( \alpha\), but the total EoS remains zero, producing  matter domination. This demonstrates that although the model gives a matter-dominated solution, the interaction remains non-zero.  Model I can produce a stable (unstable) solution for $\beta>0$ $(\beta<0)$. It is seen that a positive $\beta$ cannot adequately characterize the observed (late time) cosmology. Similarly, in Model II, the derivative of \(z'\) with respect to \(z \) (of Eq.~\eqref{z_prime_cf2}) at the fixed point gives:
   \begin{equation}\label{}
   	f'(z) = -3 \beta  (\omega +1)-3 \beta  (\omega +1) \log |z|\,.
   \end{equation} 
   Near to the critical point, \(z = 0\), the derivative diverges and the point becomes unstable, whereas, \( f'(z=1) = -3 \beta  (1+\omega ) \),  
   showing that the system stabilizes(destabilize) for \(\beta>0(<0)\) as \( N \rightarrow + \infty\).  Our analysis demonstrates that the interaction between curvature and fluid alone cannot produce an accelerating solution similar to dark energy, in the simplest workable models. We do not have any hint that complicating the interaction will produce an accelerated late time cosmic expansion. Since the variable $z$ is not constrained, it is also possible to obtain critical points at infinity. We have explicitly demonstrated this in Appendix \ref{appen:cric_infinity} and found that the system does not exhibit any stable accelerating fixed point at infinity. To obtain a late time accelerated expansion phase we require the quintessence field. The total system will then have the DM sector non-minimally coupled (NMC) to gravity and the DE sector minimally coupled to gravity.  
}  
\section{Cosmology with a NMC Fluid system and a minimally coupled Quintessence scalar field}
\label{sec:quit_minimally}

In the previous section, we briefly went through the NMC fluid system's dynamics, which is inefficient in producing an accelerating solution. As a result, we shall add a Quintessence scalar field minimally to the existing system to understand the non-minimal effect in the dynamics of the overall system. One can also take a perfect fluid; however, the field approach presents more dynamical features; thus, we choose to work with the field. The extended action is
\begin{equation}\label{}
S = S_{\rm cf} + \int d^4 x \ \sqrt{-g} \  \mathcal{L}(\phi, \p_{\mu}\phi)\,,
\end{equation}
Where the scalar field Lagrangian is given as:
\begin{equation}\label{}
\mathcal{L}_\phi =  \left[\epsilon\dfrac{1}{2} \p_\mu \phi \p^\mu \phi - V(\phi) \right]\,. 
\end{equation}
Depending on the sign of $\epsilon$, the field is a canonical scalar or quintessence field or a phantom scalar field. For $\epsilon = -1$ we have quintessence field and for $\epsilon = 1$ we have the phantom field. The field equation and energy-momentum tensor of the scalar field are given by:
\begin{equation}\label{}
\begin{split}
\epsilon \nb_\mu (\nb^\mu \phi ) + \frac{d V }{d \phi} & = 0,\\
T_\m^\phi &= -\epsilon \p_\mu \phi \p_\nu \phi + g_\m \left[\frac{\epsilon}{2} \p_\alpha \phi \p^\alpha \phi  - V(\phi)\right]\,.	
\end{split}
\end{equation}
In the background of a spatially flat FLRW metric the scalar field equation is:
\begin{equation}\label{}
- \epsilon(\ddot{\phi} + 3 H \dot{\phi}) + \frac{d V}{d \phi} = 0\,.
\end{equation}
The energy density and pressure of the field are 
\begin{equation}\label{}
\rho_\phi = - \frac{\epsilon \dot{\phi}^2}{2} + V(\phi), \quad P_\phi = - \frac{\epsilon \dot{\phi}^2}{2} - V(\phi)\,,
\end{equation}
where \( \dot{\phi} = d \phi/dt\). The modified Einstein equation becomes,
\begin{equation}\label{}
G_{\m} = \kappa^2 \left(T_{\m}^{1} + T_{\m}^{2} + T_{\m} ^{\phi} \right)\,, 
\end{equation}
where $T_{\m}^{1}\,,\,T_{\m}^{2}$ are defined in Eq.~\eqref{reT1} and \eqref{reT2}. The Friedmann equations can be expressed as: 
\begin{eqnarray}
3 H^2 & = &  \dfrac{\kappa^2 (\rho + \rho_{\phi}) }{1 - 3 \alpha \mathcal{P}_{i} - 2 \alpha f }\,,\\
\dot{H} &= & \dfrac{\kappa^2 (P_{M} + P_{\phi})}{ (-2 +  \alpha f + 3 \alpha \mathcal{P}_{i})} - \dfrac{\kappa^2 (\rho + \rho_{\rm \phi}) (1 - \alpha \mathcal{P}_{i} - 4 \alpha  (\mathcal{P}_{i} + f))}{(-1 + 2 \alpha f + 3 \alpha \mathcal{P}_{i})(-2 +  \alpha f + 3 \alpha \mathcal{P}_{i})}\,.
\end{eqnarray}
In this case, the covariant derivative of the field stress tensor conserved independently, $ \nb^{\mu}T^{\phi}_{\m} = 0 $. 

\subsection{Dynamical System Stability}

Here we present the dynamical evolution of the system we have discussed previously. Before we proceed we specify some dimensionless variables which we will use in our analysis. We shall concentrate on the canonical scalar field \( \epsilon = -1\). The standard variables chosen for this new system are:
\begin{equation}\label{eq:V1}
x = \frac{\kappa \dot{\phi}}{\sqrt{6} H}, \quad y= \frac{\kappa \sqrt{V(\phi)}}{\sqrt{3 }H}, \quad   z= f,   \quad \sigma^2 = \dfrac{\kappa^2 \rho}{3 H^2\, },\quad  \lambda = -\dfrac{V_{,\phi}}{\kappa V},\quad \Omega_{\phi} = \frac{\kappa^2\rho_\phi}{3 H^2}\,.
\end{equation}

\subsubsection{Power-law type interaction}
We choose the the interaction term as: 
\begin{equation}\label{power_law}
f(n,s)= M^{-4\beta} \rho^{\beta} (n,s)\,,
\end{equation}
where \( M\) is the mass dimension constant. This is perhaps the most simple term which can be chosen. We choose the quintessence potential as:
\begin{equation}\label{}
V(\phi)  = V_0 e^{\lambda \kappa \phi}\,.
\end{equation}
This is the standard quintessence potential used by many authors. Using these forms of interaction and scalar field potential we can now express the fluid energy density parameter as:
\begin{equation}\label{eq:f4}
\sigma = \sqrt{1-\alpha z (2+3(\beta(1+\omega)-1))-x^2-y^2}\,,	
\end{equation}
where $\omega$ is the EoS of background fluid. The 3D autonomous equations in the present case can be written as: 
\begin{subequations}
\begin{eqnarray}
x' & = & - 3 x - 3 x \left[ \frac{x^2-y^2+\omega \sigma^2}{(-2+ 3 \alpha \mathcal{P}_{i} +  \alpha z)} - \frac{(\sigma^2 + x^2 + y^2) (1-\alpha \mathcal{P}_{i} - 4\alpha (\mathcal{P}_{i} + z))}{(-1+ 3 \alpha \mathcal{P}_{i} + 2 \alpha z)(-2+ 3 \alpha \mathcal{P}_{i} +  \alpha z)}\right]\nonumber\\ 
&& + 3 \lambda y^2/\sqrt{6}\,,
\label{eq:x_prime1}\\
y' & = & - 3 y \left[ \frac{x^2-y^2+\omega \sigma^2}{(-2+ 3 \alpha \mathcal{P}_{i} +  \alpha z)} - \frac{(\sigma^2 + x^2 + y^2) (1-\alpha \mathcal{P}_{i} - 4\alpha (\mathcal{P}_{i} + z))}{(-1+ 3 \alpha \mathcal{P}_{i} + 2 \alpha z)(-2+ 3 \alpha \mathcal{P}_{i} +  \alpha z)}\right]\nonumber \\ 
&& - 	\sqrt{6} \lambda x y/2 \label{eq:y_prime1}\\
z' & = & -3 \beta z (1+\omega)\,,\label{eq:z_prime1}
\end{eqnarray}
\end{subequations}
where the $x^\prime$ stands for $\frac{1}{H}\frac{dx}{dt}$ and similarly we define $y^\prime$ and $z^\prime$. One can also represent the derivatives as $d/dN$ where $dN=Hdt$.
%
In the present case the interaction pressure term is $ \mathcal{P}_{i} =z[\beta(1+\omega)-1] $ and the total (or effective) EoS is given by:
\begin{equation}\label{}
\omega_{\rm tot} =-1 - 2 \left[ \frac{x^2-y^2+\omega \sigma^2}{(-2+ 3 \alpha \mathcal{P}_{i} +  \alpha z)} - \frac{(\sigma^2 + x^2 + y^2) (1-\alpha \mathcal{P}_{i} - 4\alpha (\mathcal{P}_{i} + z))}{(-1+ 3 \alpha \mathcal{P}_{i} + 2 \alpha z)(-2+ 3 \alpha \mathcal{P}_{i} +  \alpha z)}\right]\,.
\end{equation}
\begin{table}[t]
	\centering
	\begin{tabular}{cccccccc}
		\hline
		Points & $ x $ & $ y $ & $ z $ & $\Omega_{\phi}$& $ \sigma^2 $ & $\omega_{\rm tot}$& Stability\\
		\hline 
		\hline
		$ P_{1} $ & 0 & 0 & 0 & 0 & 1& 0 & $ \left( -\frac{3}{2},\frac{3}{2},-3 \beta \right) $ \\
		\hline 
		$ P_{2\pm} $ & $\pm 1$ & 0 & 0 & 1 & 0 & 1&  $\left(  3,-3 \beta ,\sqrt{\frac{3}{2}} \lambda +3\right)  $\\
		\hline 
		$ P_{3} $ & $ \frac{\sqrt{\frac{3}{2}}}{\lambda }$ & $ \frac{\sqrt{\frac{3}{2}}}{|\lambda| }$ & 0 & $ \frac{3}{\lambda ^2}$& $ 1-\frac{3}{\lambda ^2} $ & 0 & $ \left( -3 \beta ,-\frac{3 \left(\lambda ^2+\sqrt{24 \lambda ^2-7 \lambda ^4}\right)}{4 \lambda ^2},\frac{3 \sqrt{24 \lambda ^2-7 \lambda ^4}}{4 \lambda ^2}-\frac{3}{4} \right) $\\
		\hline 
		$ P_{4} $ & $\frac{\lambda }{\sqrt{6}}$ & $ \frac{\sqrt{6-\lambda ^2}}{\sqrt{6}}$ & 0 & $1$ & 0 & $ \frac{1}{3} \left(\lambda ^2-3\right) $ & $ \left( -3 \beta ,\frac{1}{2} \left(\lambda ^2-6\right),\lambda ^2-3\right) $\\
		\hline
	\end{tabular}
	\caption{Critical Points and their nature corresponding to minimally coupled quintessence field $\epsilon = -1$ with non-minimally coupled pressure-less fluid with power law interaction for $\omega = 0$ and general $\beta$.}
	\label{tab:critical_pts_quint-essence}
\end{table}
The fixed points corresponding to the autonomous system are given in Tab.[\ref{tab:critical_pts_quint-essence}]. We have found five critical points; for each point, the interaction variable \( z\) remains zero. These points are independent of the interaction parameter \( \beta, \alpha\) and only depend on potential parameter $\lambda$. The nature of these points are also mentioned in Tab.[\ref{tab:critical_pts_quint-essence}]. Here we briefly discuss about the various critical points.
	
\begin{itemize}
\item Point $ P_{1}\ : $ At this point, the fluid energy density dominates over field energy density while the total EoS is zero, depicting an effective matter-dominated phase. This point always shows saddle type behavior irrespective of any choice of model parameters $\lambda$ or $\beta$.

\item Points $ P_{2\mp}\ : $ At these points, the field density dominates while the total EoS is one which signifies an ultra-stiff matter phase. This point is a saddle for positive $\beta$ and becomes unstable for negative $\beta$. Cosmologically this point is not relevant or interesting.

\item Point $ P_{3}\ : $ At this point, both the field-fluid density shows non-zero contribution, while the effective EoS is always zero, which signifies that the universe near this point is in an effective matter dominated phase. The point is a saddle point for \( 0 < |\lambda| < \sqrt{3}\).

\item Point $ P_{4}: $ This point shows the dominance of field energy density over the fluid energy density. The point becomes stable when \( - \sqrt{3} < \lambda < \sqrt{3}\) and $\beta > 0 $. For the stable range, the point can exhibit the accelerating expansion phase.  
\end{itemize}
It is seen that for all the above critical points we have zero non-minimal curvature coupling. This is an interesting observation, it shows cosmological dynamics in such systems always prefer critical points where the non-minimal interaction vanishes. We will shortly see that this fact has more to do with the kind of interaction function, $f(n,s)$, we choose. 
\begin{figure}[t]
	\centering
	\includegraphics[scale=0.7]{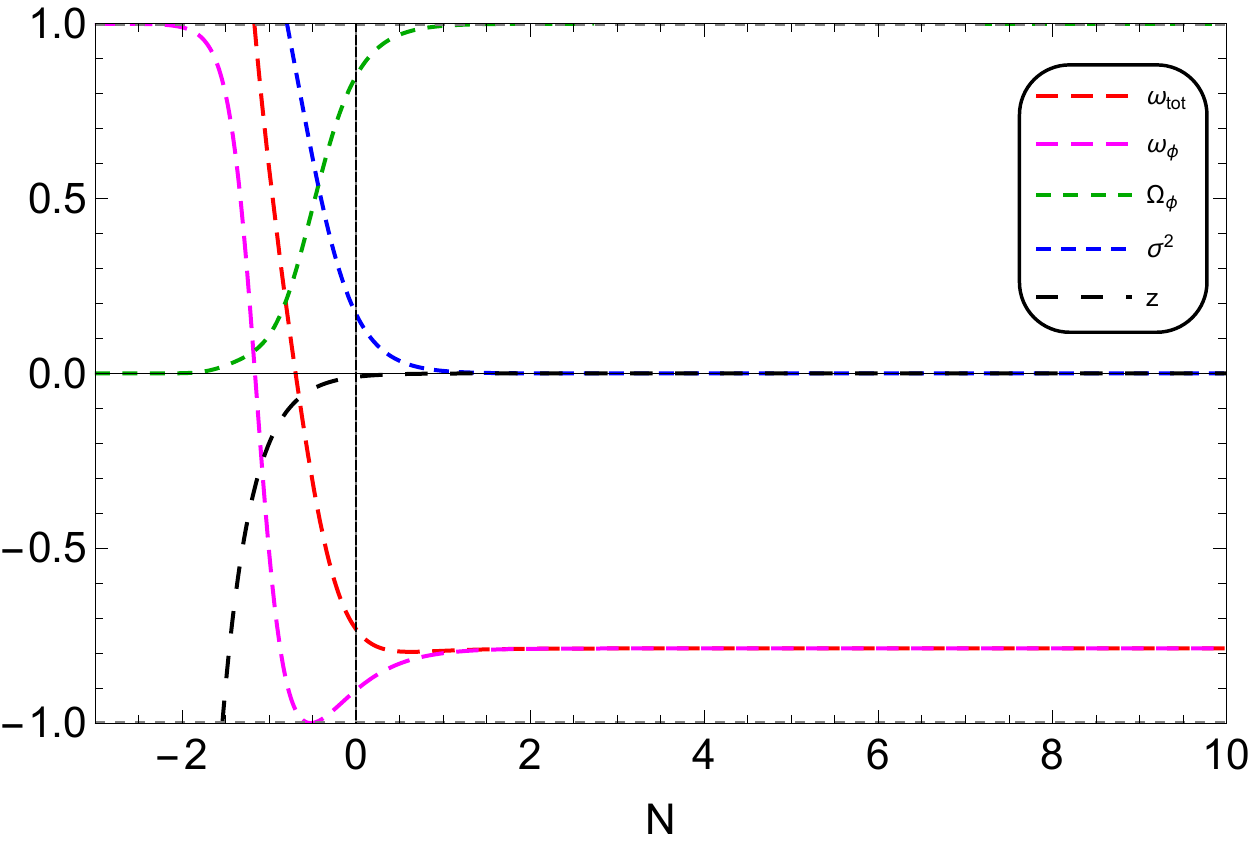}
	\caption{Evolution of the cosmological variables corresponds  to $ \lambda = 0.8, \beta= 1, \alpha = 1, \omega = 0, \epsilon =-1$. }
	\label{fig:evo_quint}
\end{figure}
The critical points deviate trivially from the minimally coupled field-fluid scenario, see Ref.\cite{Copeland:2006wr}, except that a stable accelerating solution requires the interaction parameter $\beta$ to be positive in the present case. However, the dynamical evolution of the cosmological parameters may result in some deviation from the minimally coupled field-fluid scenario. To trace the evolution of cosmological parameters, we have presented an evolutionary plot of cosmological variables against N \(= \log a \) in Fig.[\ref{fig:evo_quint}] for $\beta = 1$. The accelerating expansion point mainly depends on model parameter $ \lambda $. As  $\lambda \rightarrow 0$, total EoS parameter of this coupled system goes to $-1$. To study the complete dynamics of the system we have chosen the benchmark value of the model parameter $\lambda$ as $0.8$. During the early phase, the field and fluid energy densities are affected by the non-zero contribution of interaction parameter \( z\). In the presence of this interaction parameter,  the fluid density abruptly increases, and total EoS becomes greater than $1$. {This kind of behavior can be understood from the nature of the critical points at infinity and further discussion, regarding this matter, can be seen in Appendix \ref{power_law_infinity}.} In the early phase the field energy density tends to zero with the value of field EoS \( \omega_{\phi}\) near about  $1$. Therefore the non-zero effect of interaction parameter leads to some inconsistency and non-physical behavior in the early phase, however, at the late-time phase the interaction becomes zero. In this case we see that the only stable critical point has zero non-minimal curvature coupling. If the cosmological system evolves to such a state then the universe enters a quintessence dominated phase. The other critical points are either saddle points or unstable and the system never settles down near those points. 

As pointed out earlier we can have stable fixed points with non-minimal curvature couplings if the interaction term $f(n,s)$ is altered from the simplest possible form. In the next section, we will illustrate the case of exponential interaction which may result in non-zero curvature interaction in the late-time phase. 

\subsubsection{Exponential type interaction}

\begin{figure}[t]
	\begin{minipage}{0.30\linewidth}
		\centering
		\includegraphics[scale=0.5]{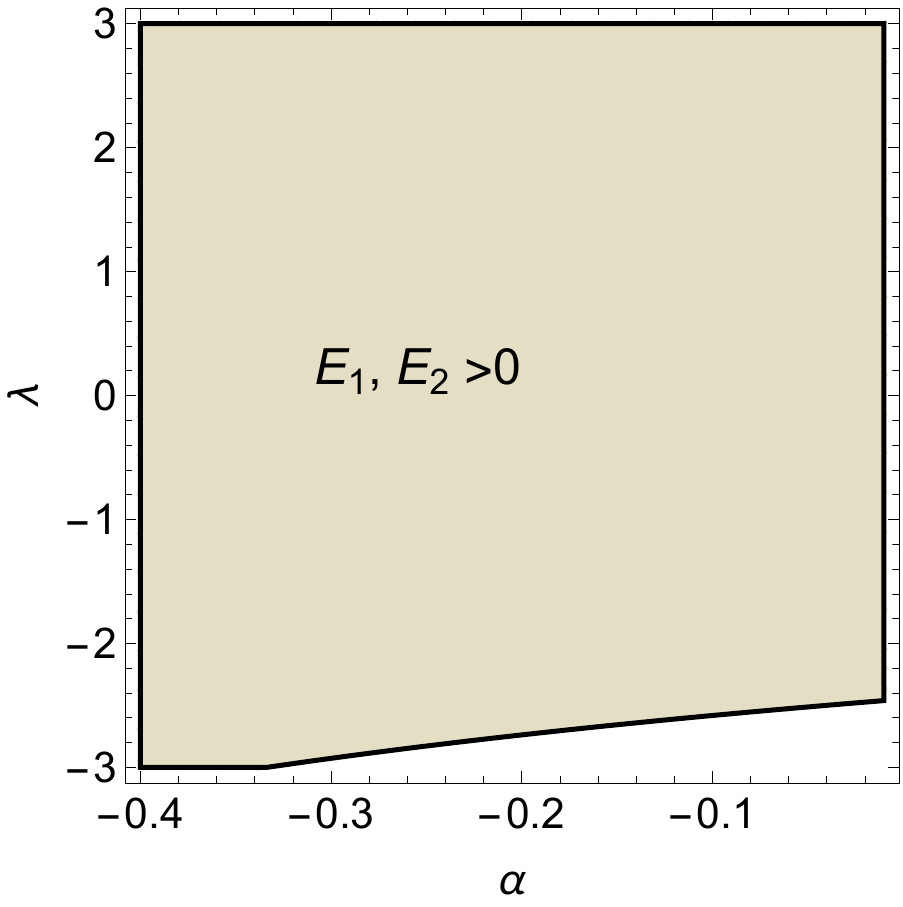}
		\caption{Stability criterion of critical point $P_{2}$.}
		\label{fig:stab_p2_min}
	\end{minipage}
		\hspace{0.2cm}
		\begin{minipage}{0.30\linewidth}
			\centering
			\includegraphics[scale=0.5]{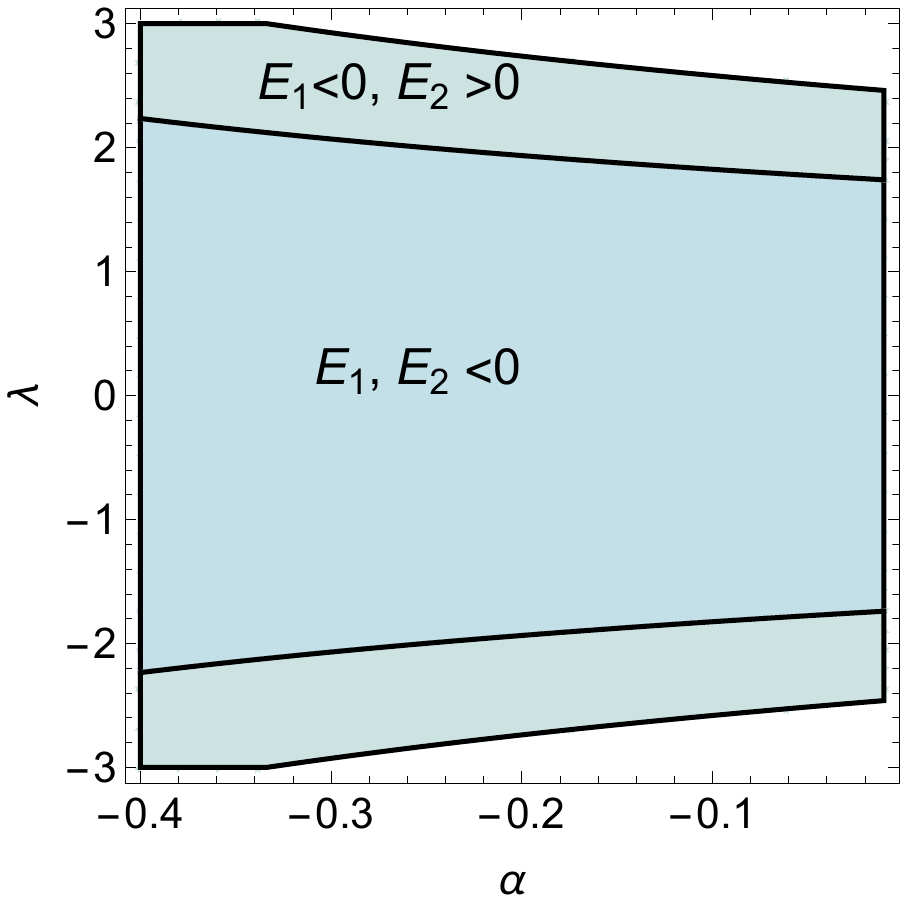}
			\caption{Stability criterion of critical point $P_{3}$.}
			\label{fig:stab_p3_min}
		\end{minipage}
		\hspace{0.2cm}
		\begin{minipage}{0.30\linewidth}
			\centering
			\includegraphics[scale=0.5]{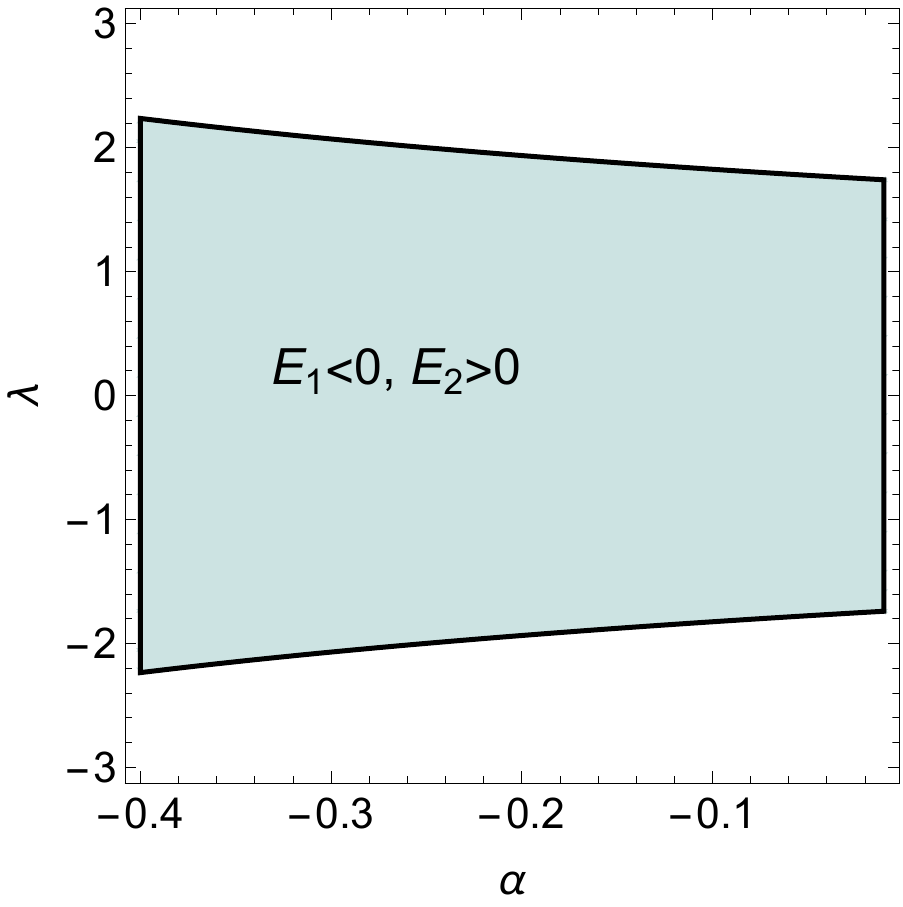}
			\caption{Stability criterion of critical point $P_{4}$.}
			\label{fig:stab_p4_min}
		\end{minipage}
\end{figure}

\begin{figure}[t]
		\begin{minipage}{0.5\linewidth}
			\centering
			\includegraphics[scale=0.6]{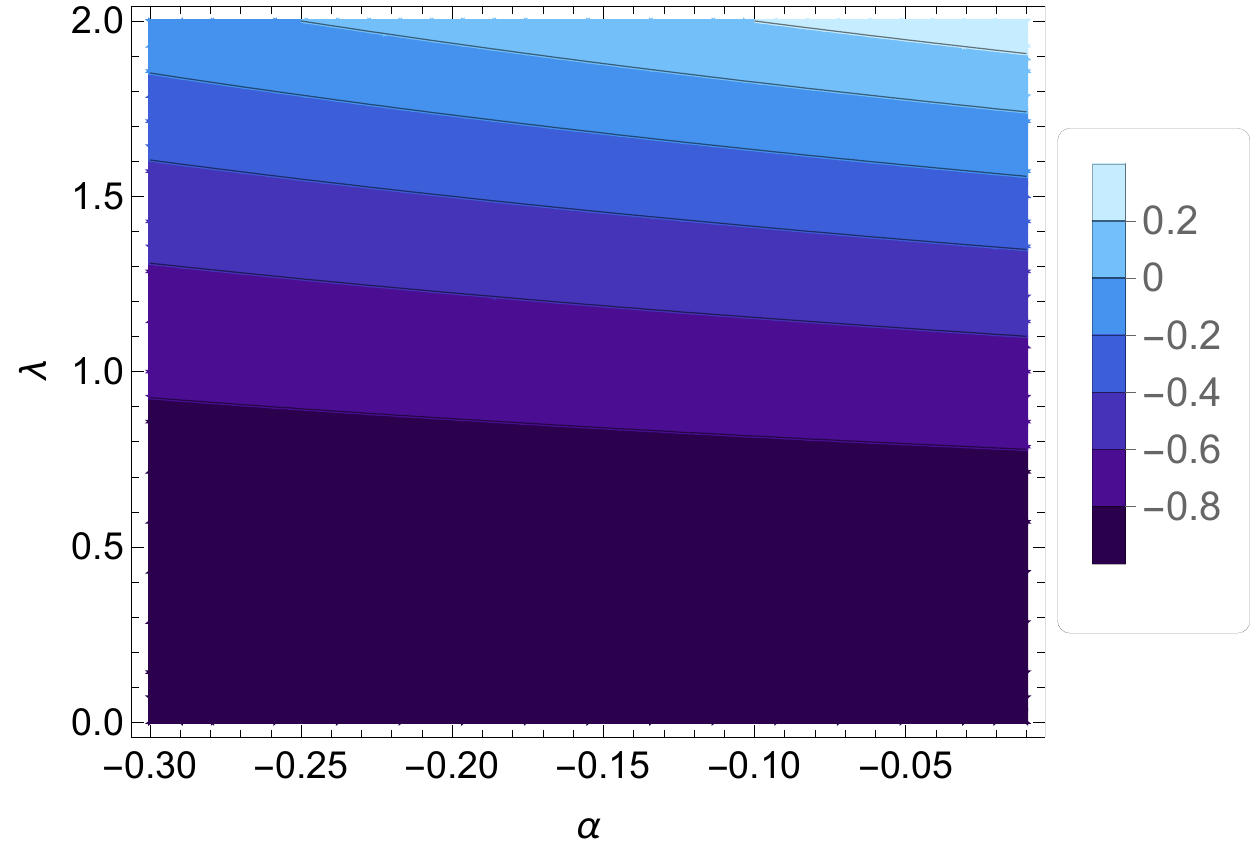}
			\caption{Various possible effective EoS of state at $P_{3}$.  }
			\label{fig:eos_expo_min}
		\end{minipage}
		\hspace{0.2cm}
		\begin{minipage}{0.5\linewidth}
			\centering
			\includegraphics[scale=0.6]{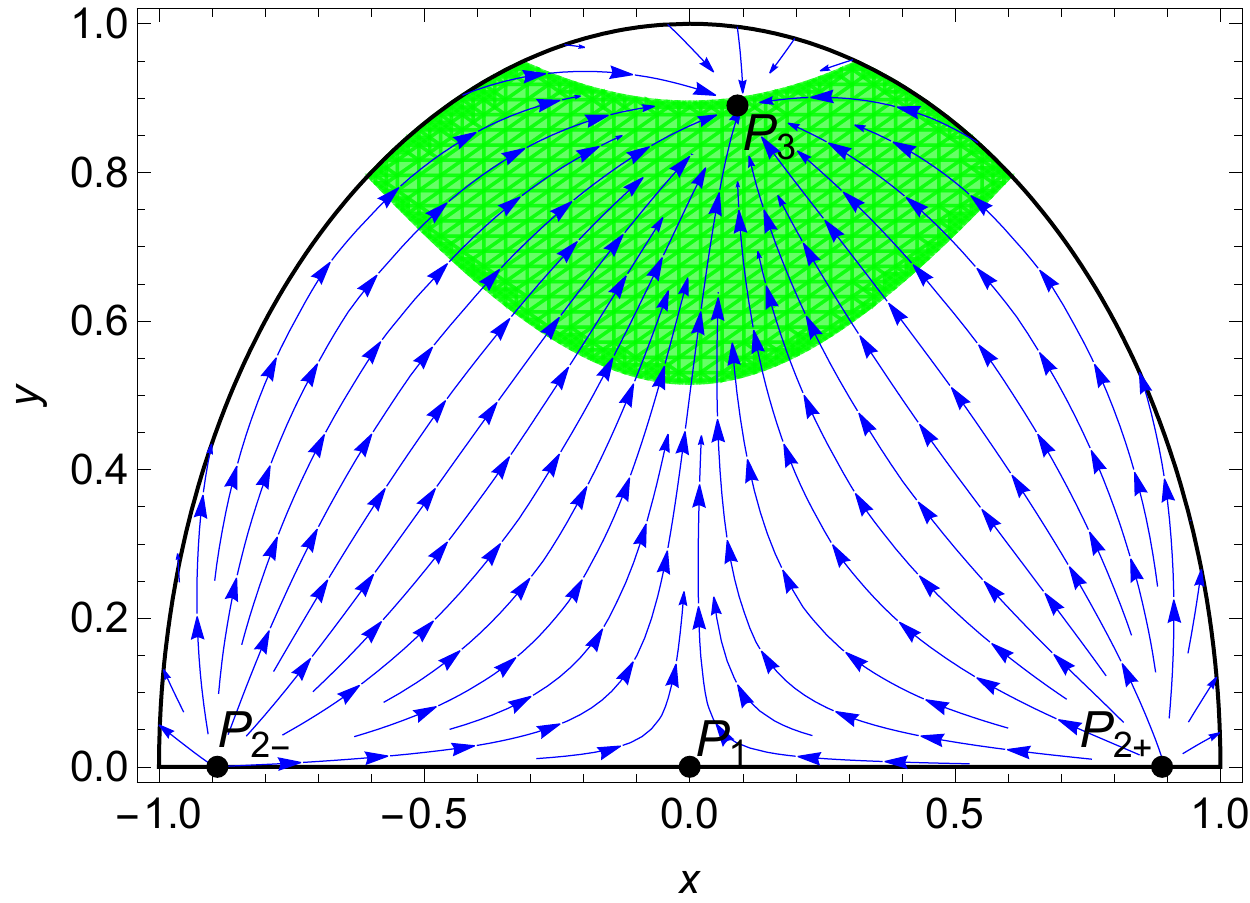}
			\caption{Phase space of exponentially interacting system in the \(z = 1\) plane for \( \alpha = -0.2, \beta = 2, \lambda = 0.3\),  {where green region shows the accelerating regime  $-1 \leq \omega_{\rm tot} \leq -1/3$.} }
			\label{fig:phase_expo_min}
		\end{minipage}
\end{figure}

\begin{figure}[t]
	\centering
	\includegraphics[scale=0.6]{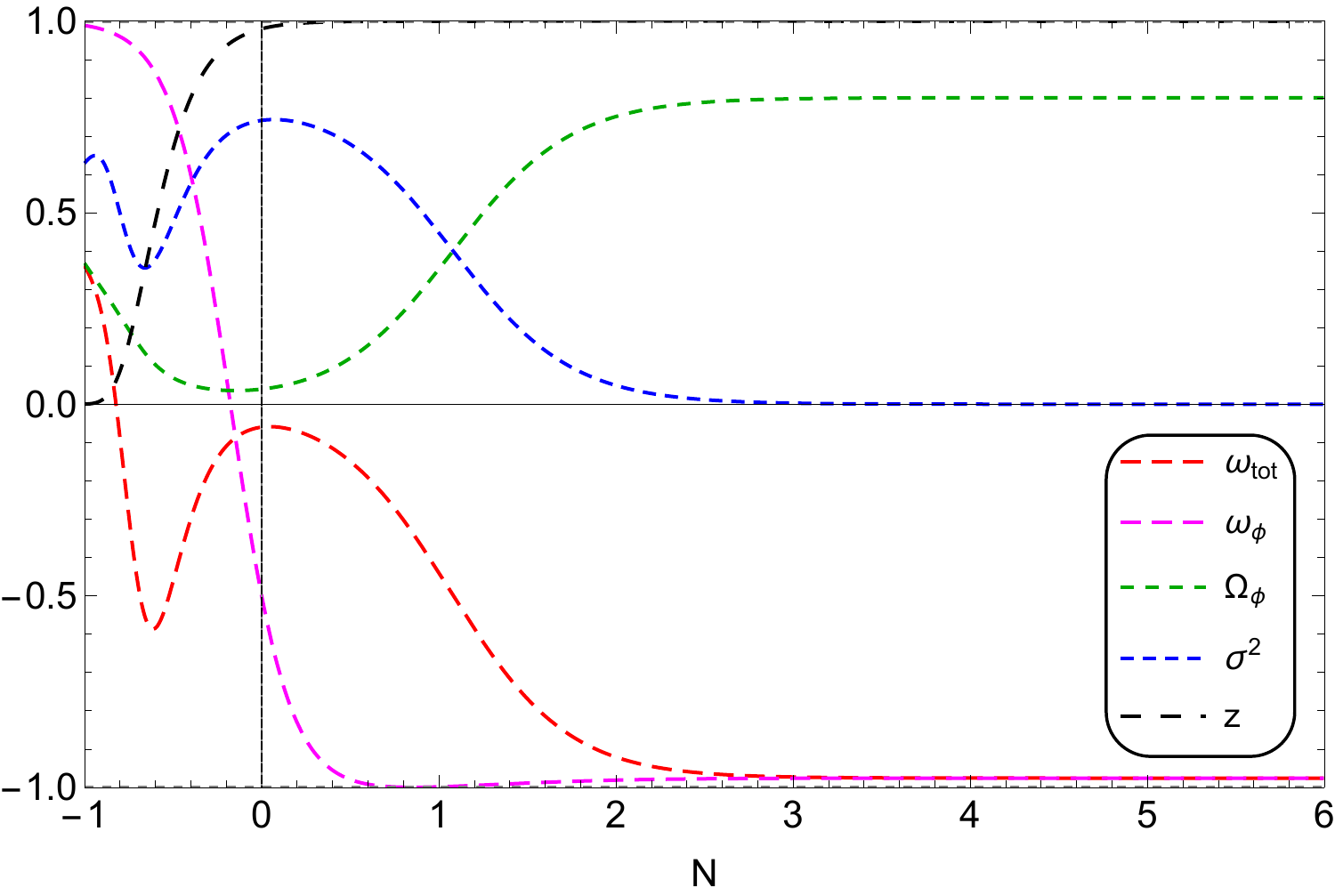}
	\caption{Numerical evolution of cosmological variables for \(\alpha = -0.2, \beta = 2, \lambda = 0.3\). }
	\label{fig:evo_expo_min}
\end{figure}

In the present case we choose
\begin{equation}
f(n,s) = \exp(M^{-4 \beta }\rho^{\beta}(n,s))\,.
\label{int_expo_quint_mini}
\end{equation}
The interaction pressure parameter and the fluid density energy density parameter are given as: 
\begin{equation}
\pp = z (\beta  (\omega +1) \ln |z| -1), \quad \sigma^2 = -x^2-y^2-3 \alpha  \beta  (\omega +1) z \ln |z|+\alpha  z+1\,.
\end{equation}
Corresponding autonomous equation is 
\begin{equation}
z' = -3 \beta z (1+\omega) \ln|z| \label{exp_z_prime}
\end{equation}
Using Eqs.~\eqref{eq:x_prime1}, \eqref{eq:y_prime1} and \eqref{exp_z_prime}, the critical points for non-zero interaction are tabulated in Tab.[\ref{tab:cric_expo_int_minimal}]. At \(z =1\), \(z'\) vanishes, therefore, all the critical points have been evaluated only for \(z = 1\). The other critical points are at \(z =0\). As because we are particularly interested in those critical points where we have nonzero curvature coupling with DM sector we only concentrate on the critical points at $z=1$ in the present case.
\begin{table}[t]
	\small
	\centering
	\begin{tabular}{ccccccc}
		\hline
		\multicolumn{7}{c}{Critical points at \(z = 1\). }\\
		\hline
		\hline
		Points & $ x $ & $ y $  & $\Omega_{\phi}$& $ \sigma^2 $ & $\omega_{\rm tot}$& Stability\\
		\hline 
		\hline
		$ P_{1} $ & 0 & 0  & 0 & 1& 0 & $ \left( -\frac{3}{2},\frac{3}{2},-3 \beta \right) $ \\
		\hline 
		$ P_{2\pm} $ & $\pm \sqrt{\alpha +1}$ & 0  & 1 & 0 & 1&  $(  -3 \beta,E_{1},E_{2} ) $ \text{Fig.[\ref{fig:stab_p2_min}]}\\
		\hline 
		$ P_{3} $ & $ \frac{(\alpha +1) \lambda }{\sqrt{6}}$ & $\frac{\sqrt{-\left((\alpha +1) \left((\alpha +1) \lambda ^2-6\right)\right)}}{\sqrt{6}}$  & $1+\alpha $ & 0 & $\frac{1}{3} (\alpha +1) \lambda ^2-1 $ & $(-3 \beta,E_{1},E_{2}  )$  \text{Fig.[\ref{fig:stab_p3_min}]}\\
		\hline 
		$ P_{4} $ & $\frac{1}{\lambda } \sqrt{\frac{3}{2}}$ & $ \frac{1}{\lambda } \sqrt{\frac{3}{2}}$ & $\frac{3}{\lambda ^2}$ & $\alpha -\frac{3}{\lambda ^2}+1$ & $0$ & $(-3 \beta ,E_{1},E_{2} )$ \text{Fig.[\ref{fig:stab_p4_min}]} \\
		\hline
	\end{tabular}
	\caption{Critical Points of minimally coupled quintessence field and non-minimally coupled pressure-less background fluid, with exponential interaction term, for \( \omega = 0,\ \epsilon = -1\) and general $\beta$.}
	\label{tab:cric_expo_int_minimal}
\end{table}
{The critical points are

\begin{itemize}
\item Point $P_1$: Both the field parameters \((x,y)\) vanish; consequently, the fluid density dominates with an effective EoS zero, indicating a matter dominated phase. The point always yields a saddle solution regardless of the choice of the model parameters. 

\item Points $P_{2\mp}$: This point denotes stiff matter solution at the early epoch owing to the dominance of the kinetic component of the field over the potential parameter, i.e., \((x>> y)\). Fig.[\ref{fig:stab_p2_min}] demonstrates that the point always exhibits saddle/unstable behavior, regardless of the value of $\beta$.

\item Point $P_3$: The characteristics of this critical point are contingent on the model parameters ($\alpha, \lambda$). The field energy density is \(\Omega_{\phi}  = 1 + \alpha\), constraining $\alpha$ to be negative. In Fig.[\ref{fig:stab_p3_min}], the distinct regions have been identified where the eigenvalues $(E_{1}, E_{2})$ are negative or pick alternating sign in the parameter space of \((\alpha, \lambda)\) for any $\beta$. The point stabilizes for positive $\beta$ and additional constraints on the model parameters can be obtained from the effective EoS $(\omega_{\rm tot})$ shown in Fig.[\ref{fig:eos_expo_min}]. The contour shows that as we decrease the value of $\lambda$, the effective EoS for this point converges towards \(-1\) for negative values of $\alpha$.  

\item Point $P_4$: This point depicts a non-accelerating solution with non-vanishing field parameters. Both the field and fluid density are finite and dependent on the gradient of the potential $\lambda$. The point in Fig.[\ref{fig:stab_p4_min}] represents only the saddle solution.

\end{itemize}
}
The phase space in Fig.[\ref{fig:phase_expo_min}] has been plotted in \(x,y\) at \(z =1\) for \(\alpha = -0.2, \beta = 2, \lambda = 0.3\). 
At fixed \(z\), the phase space gets reduced from 3-dimensions to 2-dimensions without losing any generality. The green region depicts the acceleration \(-1 \le \omega_{\rm tot} \le -1/3\). The phase space is constrained by the conditions  \(0 \le \sigma^2\le 1\) and \(0 \le \Omega_{\phi}\le 1\). The trajectories are originating from \(P_{2-}\) and \(P_{2+}\), where the field kinetic energy dominates, they are initially attracted towards \(P_{1}\) where total EoS is zero signifying a matter dominated phase. After getting repelled from \(P_{1}\), all the nearby trajectories get attracted towards \(P_{3}\) which lies in the accelerating expansion region. At these benchmark points \(P_{4}\) violates the constrained equation and becomes physically non-viable. Solving the autonomous equations for \(x',y',z'\)
numerically for these benchmark points, the cosmological observables have been plotted in Fig.[\ref{fig:evo_expo_min}]. Initially, the field EoS $\omega_{\phi}$ starts from \(1\) showing a stiff matter dominated initial phase where both field and fluid energy densities are non-zero. As number of e-folds increases the fluid density and the corresponding total EoS $\omega_{\rm tot}$ tends to zero. As the fluid density starts decaying, the field density increases and saturates at \(\Omega_{\phi} \approx 0.8\) and consequently field EoS and $\omega_{\rm tot}$ converges to \(\approx -1\). During the entire evolution, the interaction \(z\) is non-zero and at the late-time phase it saturates to \(1\). 
This shows that the quintessence field coupled minimally with NMC fluid can produce a stable acceleration during the late-time epoch and shows significant deviation from minimally coupled field-fluid scenario. {We also found a critical point at infinity in Appendix. \ref{power_law_infinity} which can not produce a stable accelerating solution during the late-time epoch}. In summary, the non-zero interaction, produces non-trivial behavior in the early phase, where both the field and fluid density contribute to stiff matter and during late-time phase, the field energy density parameter approaches a value around \(0.8\). 

\section{Simultaneous non-minimal curvature coupling of field and fluid}
\label{sec:quint_nonminimally}

In the previous section, we showed the cosmological dynamics of a system where the DM sector was non-minimally coupled to curvature whereas the DE sector was minimally coupled. Some of the solutions could produce stable late-time accelerated expansion solutions. It is natural to extend the analysis and introduce curvature coupling for the field-fluid system. Therefore, in this present section, we extend our analysis and modify the interaction term \(f(n,s) \to f(n,s, \phi)\).  The extended action is 
\begin{eqnarray}\label{action_coupled_quintessence}
S &=& \int_{\Omega} d^4 x \left[\sqrt{-g} \dfrac{R}{2 \kappa^2} -\sqrt{-g} \, \rho(n,s) +
J^\mu(\varphi_{,\mu} + s\theta_{,\mu} + \beta_A\alpha^A_{,\mu})\right.\nonumber \\
&+&\left. \sqrt{-g}\mathcal{L}(\p_{\mu} \phi, \phi)\ +\sqrt{-g}  \alpha f(n,s,\phi) \frac{R}{2\kappa^2}\right]\,. 
\end{eqnarray} 
Varying this action with respect to the field $\phi$ produces the modified Klein-Gordon equation in the background of the spatially flat FLRW metric: 
\begin{equation}\label{quint_field_eqn_nonmin}
- \epsilon(\ddot{\phi} + 3 H \dot{\phi}) + \frac{d V}{d \phi} - \frac{\p f}{\p \phi} \frac{\alpha R}{2 \kappa^2} = 0\,.
\end{equation}
Due to presence of extended functional form of the interaction term, modified Friedmann equations become:
\begin{eqnarray}
3H^2 \left(1+ \alpha f - 3 \alpha n f_{,n}\right)  + 3 \alpha H f_{,\phi } \dot{\phi} &=& \kappa^2 ( \rho_M  +  \rho_\phi)\,,\label{frd1_nonmin_quint}\\
-2\dot{H}\left(1+ \alpha f - 3 \alpha n f_{,n} \right) -3 H^2 \left(1+ \alpha f -  \alpha n f_{,n} - 9n^2 \alpha f_{,nn}\right) &-&\nonumber\\
H \left(2 \alpha f_{, \phi} \dot{\phi} - 6 \alpha n f_{, n \phi} \dot{\phi} \right) - \alpha f_{, \phi \phi} \dot{\phi}^2 - \alpha f_{, \phi} \ddot{\phi} &=& \kappa^2 (P_M  + P_\phi)\,.\label{frd2_nonmin_quint}
\end{eqnarray}
More cross terms appear in the derivative of \(f\) due to presence of an extended interaction term. We redefine the stress tensor as: 
\begin{eqnarray}\label{}
T_{\m}^1 &=& 	\left( \rho -  \dfrac{1}{2 \kappa^2} \alpha f R \right) U_{\mu} U_{\nu}   
+ \left( n \rho_{,n} - \rho - \dfrac{\alpha R}{2 \kappa^2} (n f_{,n} -f)   \right)  (U_{\mu} U_{\nu} + g_{\m} ) + T_{\m}^{\phi}\,,\\
T_{\m}^2 &=&\dfrac{-1}{ \kappa^2}\bigg[   \alpha f R_{\m} + \alpha (g_{\m} \nb_{\sigma} \nb^{\sigma} f - \nb_{\mu} \nb_{\nu} f)\bigg]\,.
\end{eqnarray}
Using Eq.~\eqref{quint_field_eqn_nonmin}, we can write the covariant derivative of the stress tensor as:
\begin{eqnarray}\label{}
&&	\nb_{\mu} T^{\m}_1 =\dfrac{-\alpha}{2\kappa^2} \left(f \nb_{0}R + f_{,\phi} \dot{\phi} R \right) + \frac{\p f}{\p \phi} \frac{\alpha R}{2 \kappa^2} \dot{\phi}\,.\nonumber\\
&&	\nb_{\mu} T^{\m}_2 =	\dfrac{-\alpha}{\kappa^2} \left[ 9 H (nf_{,n} -f) \dfrac{\ddot{a}}{a} - 3f \left( \dfrac{\dddot{a}}{a} - \dfrac{\ddot{a}}{a^2}\dot{a}\right)+ 3 H f (2 H^2 + \ddot{a}/a) - 9 \p_{0} (H^2 f_{,n } n) \right]\,.\nonumber
\end{eqnarray}
The Bianchi identity implies $ \nb_{\mu} \left(T_{1}^{\m} + T_{2}^{\m}\right) = 0 $, and hence we can write: 
\begin{equation}\label{add_fluid_non_eq}
3 n \dot{H}f_{,n}=9 n^2 H^2 f_{,nn} +  12 n H^2f_{,n}- f_{, \phi} \dot{\phi} H - 6 n Hf_{,n \phi} \dot{\phi} + f_{,\phi \phi} \dot{\phi}^2  + f_{,\phi} \ddot{\phi}\,.
\end{equation}
The above equation connects \(f_{,nn}\) with $\ddot{\phi}$. Using Eq.~\eqref{quint_field_eqn_nonmin}, \(\ddot{\phi}\) can be eliminated and hence the above condition can be linked with the evolution of the field sector. Using equation  Eq.~\eqref{frd2_nonmin_quint}, we can further modify the second Friedmann equation. The total (or effective) EoS can be expressed as:  
\begin{multline}\label{eos_nonmin_quint}
\omega_{\rm tot} = -\frac{2 \dot{H}}{3H^2} -1 = -1 + \frac{P_M + P_\phi}{\rho_M + \rho_\phi} \left[\frac{1+ \alpha f -3 \alpha n f_{,n} + 3 \alpha f_{,\phi} \phi'}{1+\alpha f - 3 \alpha n f_{,n} - 9\alpha n f_{,n}/2}\right] + \\
\bigg[\frac{1+ \alpha f + 13 \alpha n f_{,n} - \alpha f_{,\phi} \phi'/3 - 7 n \alpha f_{,n\phi} \phi' + 4\alpha f_{,\phi \phi} \phi'^2 /3 + 4\alpha f_{,\phi} \ddot{\phi}/(3 H^2)}{1+\alpha f - 3 \alpha n f_{,n} - 9\alpha n f_{,n}/2}\bigg]\,.
\end{multline}
Here \( \phi' \equiv \dot{\phi}/H\). Therefore the coupled field-fluid-curvature scenario gives rise to a complex dynamical system. We will examine this complex scenario from the  dynamical system perspective, assuming the exponential form of quintessence potential \( V(\phi) =  V_0 e^{\lambda \kappa \phi}\). 

\subsection{Dynamical Stability Analysis}

\begin{figure}[t]
		\centering
		\includegraphics[scale=0.6]{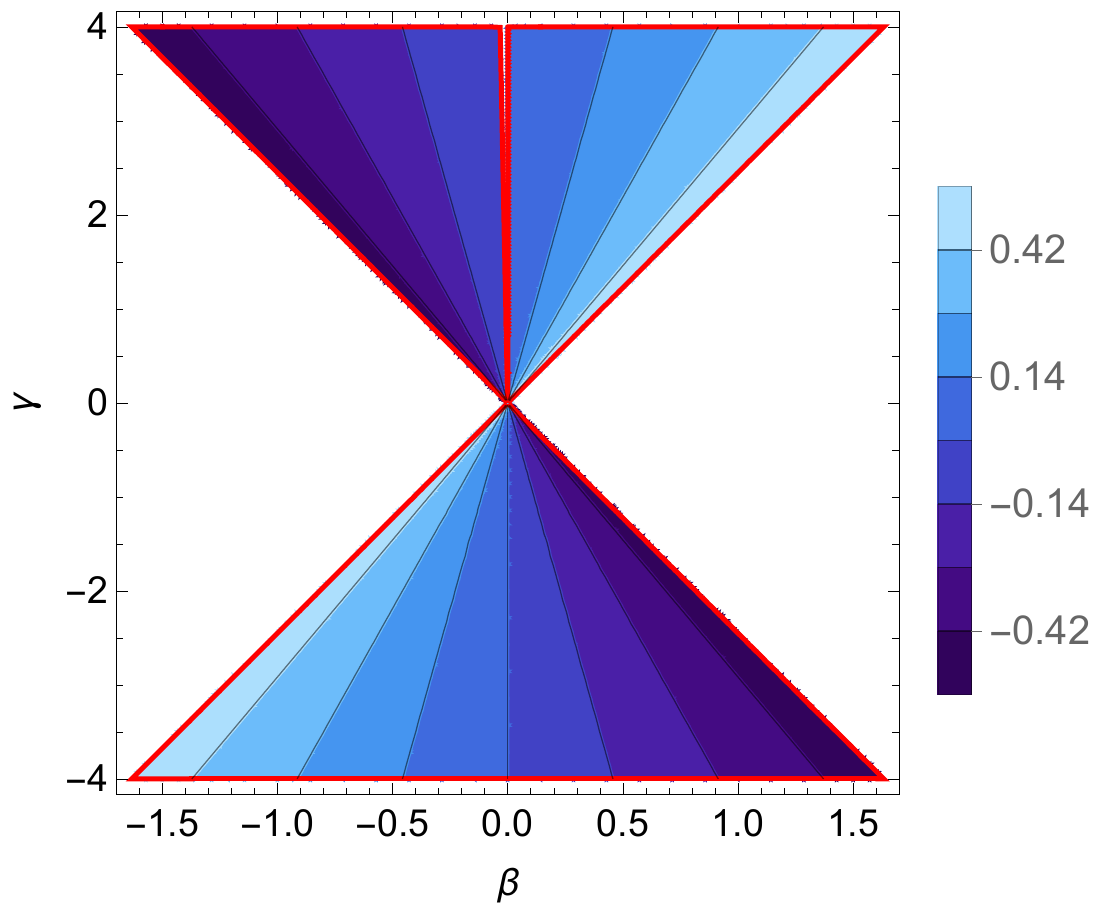}
		\caption{The possible values of $x= \frac{1}{\gamma}{\sqrt{\frac{3}{2}} \  \beta } $  in the range of \(-0.5<x <+0.5\), for various values of the parameters in the parameter space \((\beta, \lambda)\). }
		\label{fig:x_range}
	\end{figure}

\begin{figure}[t]
	\begin{minipage}{0.5\linewidth}
		\centering
		\includegraphics[scale=0.74]{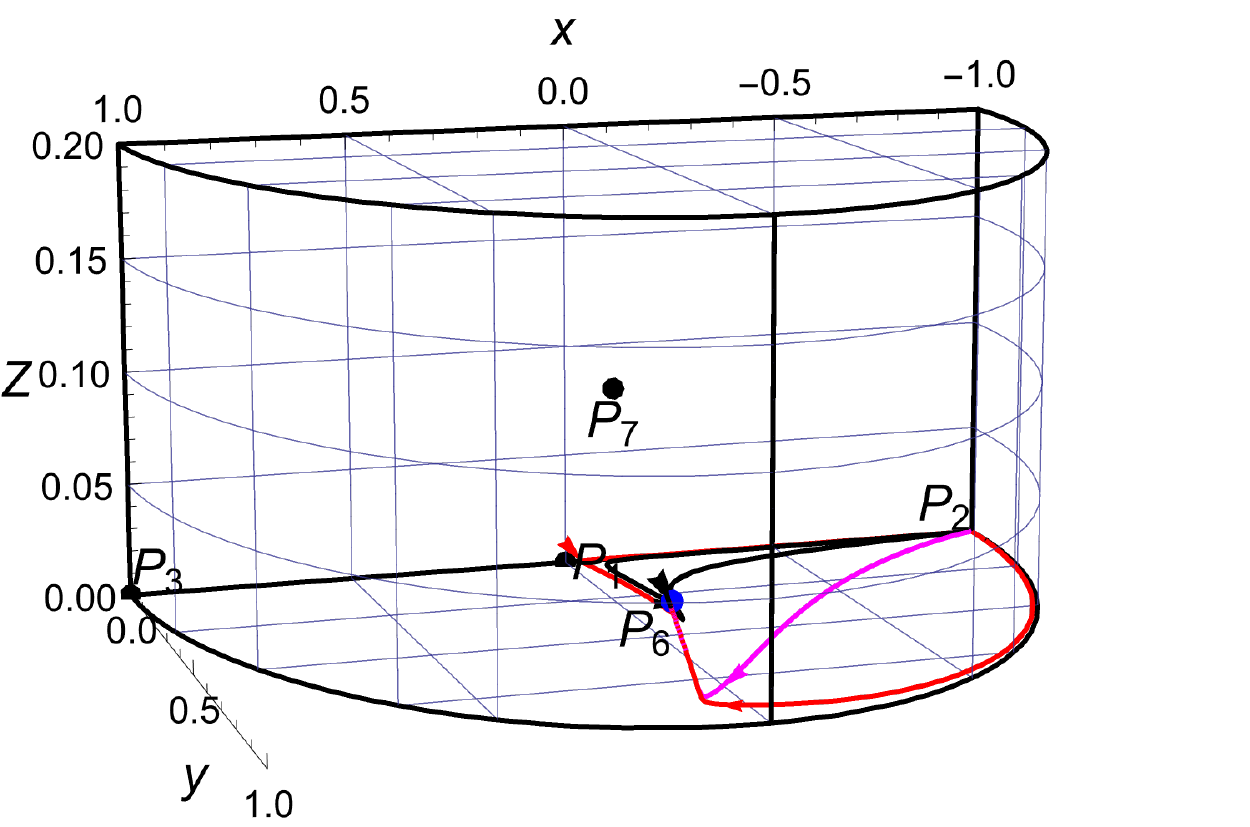}
		\caption{The phase space trajectory for $ \beta=-0.3, \gamma = -1.9, \alpha =2.0, \lambda = -0.3,\epsilon = -1  $.}
		\label{fig:phase_Quint_bet03}
	\end{minipage}
	\hspace{0.3cm}
	\begin{minipage}{0.5\linewidth}
		\centering
		\includegraphics[scale=0.6]{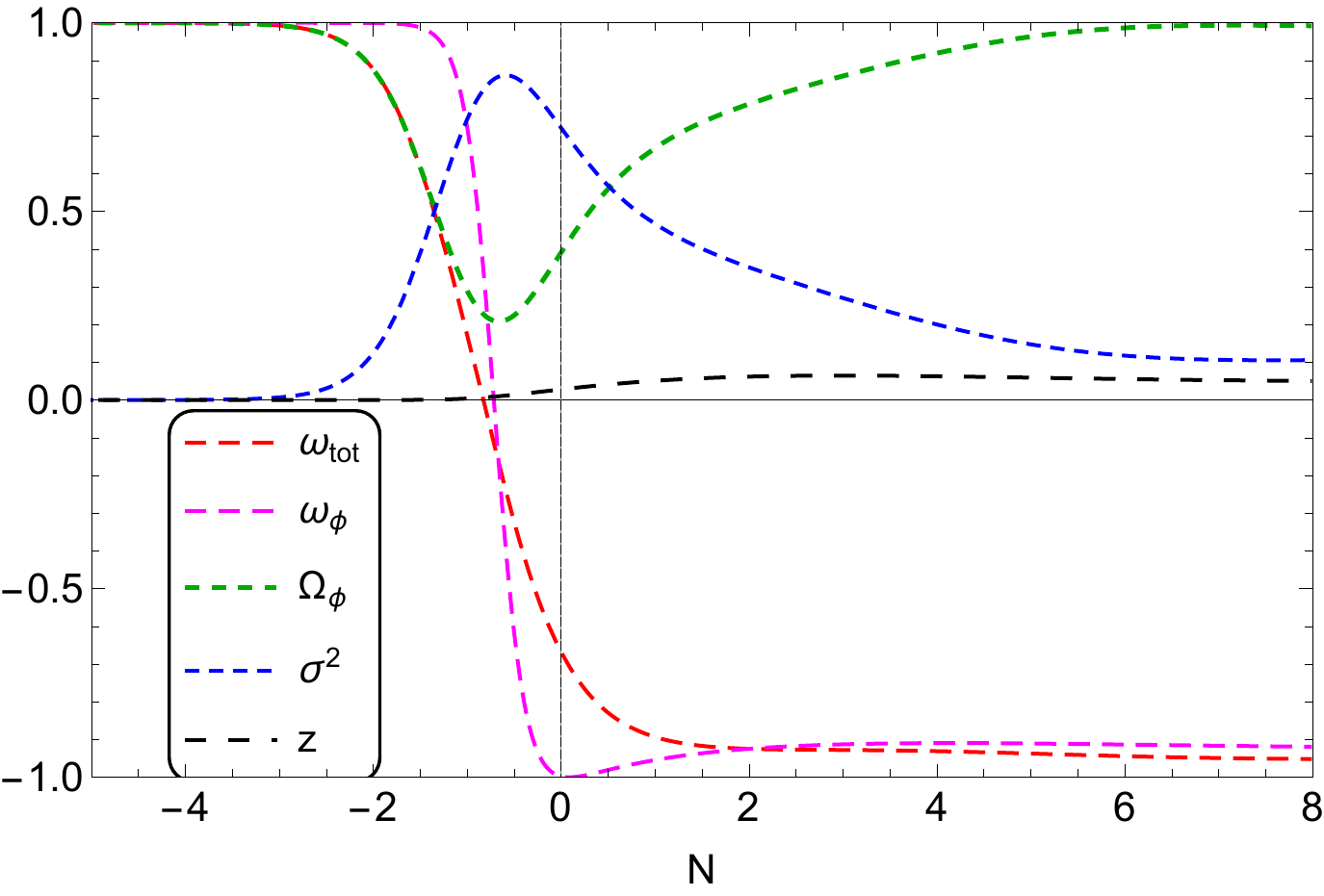}
		\caption{The evolution of the cosmological variables corresponding to $ \beta=-0.3, \gamma = -1.9, \alpha =2.0, \lambda = -0.3,\epsilon = -1  $.}
		\label{fig:evo_nonmin_quint_bet03}
	\end{minipage}
\end{figure}

The dimensionless variables to close the system remain the same as before, except the $\lambda$ definition has been altered a bit. The relevant dynamical variables are:
\begin{equation}\label{}
x = \frac{\kappa \dot{\phi}}{\sqrt{6} H}, \quad y= \frac{\kappa \sqrt{V(\phi)}}{\sqrt{3 }H}, \quad \sigma^2 = \frac{\kappa^2 \rho}{3 H^2 }, \quad f= z, \quad \lambda = \frac{V_{, \phi}}{\kappa V(\phi)}\,. 
\end{equation}
The Friedmann equation in Eq.~\eqref{frd1_nonmin_quint} can be written as: 
\begin{equation}\label{frd_1_non_dim}
1+ \alpha f - 3 \alpha n f_{,n}  + \frac{\alpha f_{, \phi} \sqrt{6 } x}{\kappa} = \sigma^2 - \epsilon x^2 + y^2\,.
\end{equation}
In the above equation, we see that there appears a derivatives of \( f\), hence in order to proceed further we must need to choose some model. One can construct a simplest model of the following type: 
\begin{equation}\label{}
f(n,s,\phi) =  \rho(n,s) \xi(\phi)\,.
\end{equation}
Although one can choose a wide variety of interacting models, 
all the choices of the model can't be expressed using the above-defined variables. This results in defining new variables as a result of which the dimensionality of the autonomous system increases. For example, considering the interaction to be \( f =  \rho^\beta ( \kappa \phi)^\gamma \). In this case, \( f_{, \phi} =  \gamma f / \phi\). As a result of this the Friedmann equation \eqref{frd_1_non_dim} will get \( \kappa \phi \) dependence. This factor ($\kappa \phi$)  cannot be expressed in terms of above-defined variables for the chosen form of potential \( V(\phi) = V_0 e^{\lambda \kappa \phi}\).  This factor will increase the dimension of the autonomous equations, and hence the phase space will become \( 4-\)dimension. One can also study the generalized interaction model by assuming the potential form as \( V(\phi)= V_0 e^{\lambda (\kappa \phi)^{n}}\).  In this case, the first derivative of the potential becomes $\phi$ dependent, and hence $\phi$ can be inverted in the first derivative of the potential involved in \( \lambda\). This technique has been demonstrated in Ref.\cite{Das:2019ixt} where the authors worked with minimally coupled quintessence. Keeping the restriction on the autonomous equation and working with the standard form of the potential as \( V(\phi) = V_0 e^{\lambda \kappa \phi}\), we will choose the interaction model as: 
\begin{equation}\label{quint_inter_model}
f= M^{-4\beta} \rho^{\beta} e^{\gamma \kappa \phi}\,.
\end{equation}
This choice produces the simplest dynamical system with 3-dimensional phase space.
The exponential type of interaction has been studied in Ref.\cite{Boehmer:2015kta}. Here \( M\) is a constant with mass dimension and \( \gamma, \beta \) are the  model parameters. With this interaction one can express the derivative of the Hubble parameter and the constrained equation as: 
\begin{multline}\label{hdot_non_quint}
-\frac{2 \dot{H}}{3 H^2} = \frac{\omega \sigma^2-\epsilon x^2 - y^2 }{\sigma^2 - \epsilon x^2 + y^2} \left[\frac{1+ \alpha z + 3 \alpha \beta (\omega + 1) z + 3 \sqrt{6} \alpha \gamma z x}{1 + \alpha z - 15 \alpha \beta (\omega + 1) z/2 - 6 \alpha^2 \gamma  y z^2 / \epsilon}\right]+ \\
		\frac{1+ \alpha z + 13 \alpha \beta (\omega +1) z - \sqrt{6} \alpha \gamma z x /3 - 7 \alpha \beta (\omega + 1 ) \gamma z \sqrt{6} x + 8 \alpha \gamma^2 x^2 z }{1 + \alpha z - 15 \alpha \beta (\omega + 1) z/2 - 6 \alpha^2 \gamma  y z^2 / \epsilon}\\
		+	\frac{- 8 \alpha^2 \gamma y z^2/\epsilon + 4 \alpha \gamma z y^2 \lambda / \epsilon - 4 \sqrt{6} \alpha x \gamma z }{1 + \alpha z - 15 \alpha \beta (\omega + 1) z/2 - 6 \alpha^2 \gamma  y z^2 / \epsilon}\,,
\end{multline}
and 
\begin{equation}\label{sigma_quint_non}
\sigma^2 = 1+ (\alpha z - 3 \alpha \beta (\omega + 1) z + \alpha \gamma \sqrt{6} x z) + \epsilon x^2 - y^2\,.
\end{equation}
In general the phase space is not simple in our case as because of the non-minimal coupling and the natural constraints on the energy density parameters do not materialize naturally. To make things manageable we apply some constraints on the phase space and only concentrate on the region of constrained space. In this work we will concentrate on those regions of phase space where the following conditions hold 
\begin{equation}
0\le \Omega_{\phi} \le 1, \quad 0\le \sigma^2 \le 1\,.
\label{constraint_rel}
\end{equation}
The autonomous equations in the present case are: 
\begin{eqnarray}
x^\prime & =& \frac{-2 \dot{H}}{3 H^2} \frac{3}{2} \left[\frac{3 \alpha y z }{\sqrt{6 } \epsilon} + x\right] + \frac{3 }{\sqrt{6 }} \left[\frac{-2 \alpha y z }{\epsilon} + \frac{y^2 \lambda}{\epsilon} - \sqrt{ 6} x\right]\,, \label{x_prime_non_quint}\\
y^\prime &=& \frac{\sqrt{6} x y \lambda }{2} + \frac{3y}{2} \left(\frac{- 2 \dot{H}}{3 H^2}\right)\,,\label{y_prime_non_quint}\\
z^\prime &=& -3 \beta (\omega + 1) z + \sqrt{6} \gamma z  x \label{z_prime_non_quint}\,.
\end{eqnarray}
In this autonomous equation, one can see that the interaction term \(z'\) has now coupled with field variables. Out of these three autonomous equations we see that \( y'=0\) at \(y=0\) and \( z'=0\) for \(z=0\). This implies in a 3-dimensional phase space, no phase trajectories originate with \(+y\) or \(+z\) can cross \(y=0, z=0\) line.
\begin{table}[t]
		\centering
	
		\begin{tabular}{cccccccc}
			\hline
			Points & \(x\) & \( y\) & \(z\)& $\Omega_{\phi}$ & $\sigma^2$ & $\omega_{\rm tot}$ & Eigenvalues\\
			\hline
			\hline
			\(P_1\) & 0 & 0 & 0& 0 & 1 & 0 & $ (-1.5,1.5,-3\beta) $ \\
			\hline
			\( P_{2,3}\) & $\mp 1 $ & 0 & 0 & 1 & 0 & 1 & $ \left(3,\sqrt{6} \gamma -3 \beta ,\sqrt{\frac{3}{2}} \lambda +3 \right)  $\\
			\hline
			\( P_{4}\) & $ -\frac{\sqrt{\frac{3}{2}}}{\lambda } $ &$ \frac{\sqrt{\frac{3}{2}}}{|\lambda |} $ & 0 & $ \frac{3}{\lambda ^2} $ & $ 1-\frac{3}{\lambda ^2} $ & 0 & $ \left(-\frac{3 (\beta  \lambda +\gamma )}{\lambda },-\frac{3 \left(\lambda ^3+\sqrt{24 \lambda ^4-7 \lambda ^6}\right)}{4 \lambda ^3},\frac{3 \sqrt{24 \lambda ^4-7 \lambda ^6}}{4 \lambda ^3}-\frac{3}{4} \right)  $\\
			\hline
			$ P_{5} $ & $ -\frac{\lambda }{\sqrt{6}} $ & $ \sqrt{1-\frac{\lambda ^2}{6}} $& 0 & 1 & 0 & $ \frac{\lambda ^2}{3}-1 $ & $ \left( -3 \beta -\gamma  \lambda ,\frac{1}{2} \left(\lambda ^2-6\right),\lambda ^2-3\right)  $\\ 
			\hline
			$ P_{6} $ & $ \dfrac{3 \beta}{\sqrt{6}\gamma} $ & $-$ & $ -$	& $ -$& $ -$ & $-$ & $--$	\\
			\hline
			
		\end{tabular}
		\caption{Critical points corresponding to the interaction term $ f = M^{-4 \beta } \rho^{\beta} e^{\gamma \kappa \phi} $ for the quintessence field for which $\epsilon = -1$. For a full explanation of the entries in the table consult the text.}
		\label{tab:critc_nonmin_quint}
	\end{table}
We found six critical points in this system, as shown in Tab.[\ref{tab:critc_nonmin_quint}] out of which few have been obtained by setting \(z=0\). Finding the critical points in terms of model parameters for \(z \ne 0\) is a challenging task due to the complexity of the system. Here one needs to heavily rely on the numerical technique to obtain the non-trivial critical points. Obtaining the critical points by numerical technique requires the values of all the parameters \(\alpha, \beta, \gamma, \ \& \ \lambda\). However, from the previous analysis, we understand that the crucial parameters that can severely affect the critical points are the model parameters of interaction \(\beta, \gamma\). To constrain these parameters corresponds to \(z\ne 0\), we set \(z' = 0\) in  Eq.~\eqref{z_prime_non_quint} and get
\begin{equation}
x = \dfrac{1}{\gamma}{\sqrt{\frac{3}{2}} \  \beta }\,.
\end{equation}
This condition guarantees that for \(z \ne 0\), this should be one of the coordinates of the critical points (in 3-dimensions). Just inserting this value of $x$ on other equations do not simplify the situation. It is known that for accelerated expansion at the critical point, the field energy density parameter $\Omega_{\phi} = x^2  + y^2 \approx 1$, and the scalar field potential term must dominate  the kinetic term, that implies \(y \gg x\). Following these arguments we plotted a range of \(-0.5<x< 0.5\) in the parameter space of \((\beta, \gamma)\) in Fig.[\ref{fig:x_range}] which produces a constraint on \((\beta, \gamma)\) values. We qualitatively discuss about the nature of the critical points below. 
\begin{itemize}

\item Point $P_{1-5}\ :$ These fixed points are independent of interaction parameters. The stability of the fixed points can be determined from the model parameter-dependent eigenvalues. Fixed point $ P_{1} $ is always saddle; however, the points $ P_{2,3} $ can act as a saddle or unstable fixed points based on the range of the model parameters.  Condition for getting stable $ P_{4} $ point is $ \sqrt{3}<|\lambda| \leq 2 \sqrt{\frac{6}{7}} $, $ \beta >-\frac{\gamma }{\lambda },\  \gamma \in \mathbb{R}$, otherwise this point becomes a saddle point. The point $ P_{5} $ becomes stable for $ -\sqrt{3}<\lambda <\sqrt{3}, \beta >-\frac{1}{3} (\gamma  \lambda ),\ \gamma \in \mathbb{R} $.
		
\item Point $P_{6}\ :$ As explained above, if $z\ne 0 $ and \(z' = 0\) then  Eq.~\eqref{z_prime_non_quint} gives one of the coordinates of the critical points. Getting other coordinates of the critical points $(y,z)$ in terms of the model parameters is a difficult task as there may be many of them. Therefore, following the constraints on \((\beta,  \&\,  \gamma)\) in Fig.[\ref{fig:x_range}], we evaluated critical point $P_{6}$ for several combinations of model parameters in Tab.[\ref{tab:cric_nonmin_numeric}]. We have selected those values of model parameters that exhibit an accelerating expansion phase. Some of these solutions are like phantom solutions although we are working with a purely canonical quintessence like scalar field. This phantom behavior originates from the non-minimal coupling term. We have concentrated on the region of phase space where $0 \le \Omega_{\phi}\le 1, 0\le \sigma^2 \le 1$. We find that several combinations of $\beta$ produce saddle points where the system is exhibiting accelerated expansion solution, while for values of as $(\beta =0.2, 1.0, 0.4 )$ we get repeller phantom solutions. Although we have found a stable accelerating solution for negative fractional values and positive integer values of $\beta$ yet we could not find any stable phantom solution. Nevertheless, we do not stress that stable phantom solution can not be determined, since the parameter space is extremely diverse and complexity of equations prevents us from pinpointing the critical points corresponding to the phantom solutions.  We think that on analyzing the model with the cosmological data may help to put several constraints on the parameters which may produce stable phantom solutions. 
\end{itemize}
\begin{table}[t]
		\centering
		\begin{tabular}{c|cccccc}
			\hline
			Points & $(\beta, \ \gamma, \ \lambda, \  \alpha)$ & \((x,y,z)\) & $\Omega_{\phi}$ & $\omega_{\rm tot}$& $\sigma^2$ & Stability\\
			\hline
			\hline
	\multirow{10}{1em}{$P_{6}$} & $(0.2,1.0, 0.1, 2.0)$ & \((0.25, 0.65, -0.25)\) &$0.48$ & $-1.02$ &$0.03$ &Saddle\\
	\cline{2-7}
	& \((0.4,-1.5,0.5,2.0)\) & $(-0.33,0.71,-0.19)$ & $0.61$ & $-0.87$ & $0.003$ & Saddle\\
	\cline{2-7}
	& $(-0.2,-2,-0.9,2.0)$ & $(0.12,0.91,0.14)$ & $0.85$ & $-0.91$ & $0.42$ & Saddle \\
	
	\cline{2-7}
	 & $(-0.3, -1.9,-0.3,2.0)$ & $(0.19,0.97,0.096)$ &$0.98$ & $-0.95$ & $0.12$ & Stable\\
	 \cline{2-7}
	 & $(-0.4, -3.0,-0.3,2.0)$ & $(0.16,0.94,0.034)$ &$0.91$ & $-0.96$ & $0.15$ & Stable\\
	 \cline{2-7}
	 & $(0.4,-5,-0.5,2.0)$ & $(-0.09,0.56,-0.17)$ & $0.33$ & $-1.04$ & $0.32$ & Unstable \\
	 \cline{2-7}
	  & $(-1.0,8.0,0.5,2.0)$ & $(-0.15,0.88,0.01)$ & $0.80$ & $-0.94$ & $0.21$ & Saddle \\
	 \cline{2-7}
	  & $(1.0,8.0,0.4,2.0)$ & $(0.15,0.79,0.19)$ & $0.64$ & $-1.05$ & $0.75$ & Saddle \\
	  \cline{2-7}
	  & $(1.0,-6.0,0.8,2.0)$ & $(-0.20,0.84,0.03)$ & $0.74$ & $-0.87$ & $0.32$ & Stable \\
	 \cline{2-7}
	 & $(2.0,-8.0,1.01,2.0)$ & $(-0.31,0.81,0.004)$ & $0.75$ & $-0.75$ & $0.26$ & Stable \\
	\hline
	\hline
	\end{tabular}
	\caption{Numerical values of the cosmological variables corresponding to $P_{6}$.}
	\label{tab:cric_nonmin_numeric}
	\end{table}
{
Based on the stability of the critical point \(P_{6}\) reported in Tab.[\ref{tab:critc_nonmin_quint}], we will analyze the interacting system only for specific choices of the model parameters ($\beta,\gamma,  \lambda, \alpha$). We categorize our analysis based on the values of $\beta$, which corresponds to the fluid energy density. In case I, we choose negative $\beta$ such that the interaction form becomes \((f \propto 1/\rho)\). We choose so that as the fluid density $\sigma^2$ dilutes, the interaction increases during the late-time epoch. In the other case we will examine \((f \propto \rho) \). In both cases, we shall report the new and previously discovered critical points of  Tab.[\ref{tab:critc_nonmin_quint}] in Tab.[\ref{tab:cric_case_wise_nonmin}], which satisfy the energy constraint relations given in Eq.~\eqref{constraint_rel}. In these two instances, we will not elaborate explicitly on the physical characteristics of the points already mentioned in Tab.[\ref{tab:critc_nonmin_quint}]. We will only discuss those critical points which are new and relevant. Those critical points in Tab.[\ref{tab:critc_nonmin_quint}] that are model dependent and do not exist for the selected benchmark points have been skipped. \\
\begin{table}[t]
	\centering
	\begin{tabular}{cccccc}
		\hline
		Points & $(x,y,z)$ & $\Omega_{\phi}$ & $\omega_{\rm tot}$ & $\sigma^2$ & Stability\\
		\hline
		\hline
		\multicolumn{6}{c}{Case I: $\beta = -0.3, \gamma = -1.9, \lambda = -0.3, \alpha = 2.0$ }\\
		\hline 
		\hline
		$P_{1}$ & $(0,0,0)$ & 0 & 0 & 1 & Saddle \\
		\hline
		$P_{2,3}$ & $(\mp 1,0,0)$ & 1 & 1 & 0 & Saddle\\
		\hline
		$P_{6}$ & $(0.19,0.97,0.05)$ & $0.98$ & $-0.95$ & $0.12$ & Stable\\
		\hline
		$P_{7}$ & $(0.19,0.70,0.12)$ & $0.53$ & $-0.95$ & $0.70$ & Saddle\\
		\hline
		\hline 
		\multicolumn{6}{c}{Case II: $\beta = 1, \gamma = -6, \lambda = 0.8, \alpha = 2.0$ }\\
		\hline 
		\hline
		$P_{1}$ & $(0,0,0)$ & 0 & 0 & 1 & Saddle \\
		\hline
		$P_{2}$ & $(-1,0,0)$ & 1 & 1 & 0 & Saddle\\
		\hline
		$P_{6}$ &  $(-0.20,0.84,0.03)$ & $0.74$ & $-0.87$ & $0.32$ & Stable \\
		\hline
		\hline
	\end{tabular}
	\caption{Critical points for different model parameters. }
	\label{tab:cric_case_wise_nonmin}
\end{table}

\noindent \textbf{Case I: $(\beta = -0.3, \gamma = -1.9, \lambda = -0.3, \alpha = 2.0)$} For this choice of benchmark points, the model renders five critical points in Tab.[\ref{tab:cric_case_wise_nonmin}] with \(P_{7}\) being the newest point. 
\begin{itemize}

\item  Point $P_6$: With $\omega_{\rm tot} \sim -1$, the point exhibits stable accelerating solution. The field energy density dominates the fluid energy density during the late-time and the fluid density saturates at \(\sigma^2 \sim 0.12\). Thus, the point specifies dark energy domination. 

\item Point $P_7$: The point also generates an accelerating solution, however, it appears that the acceleration is fluid driven i.e. \((\sigma^2 > \Omega_{\phi})\) and turns out to be saddle. Therefore, this point does not depict the late-time characteristics of the present universe. 

\end{itemize} 

}
The phase space is 3-dimensional and the full dynamics in this space is shown in Fig.[\ref{fig:phase_Quint_bet03}]. The variables \((x, y)\) are constrained via the relation \(0 \le x^2 + y^2 \le 1\), although the variable $z$ can take any value from \(0\) to $\infty$. Therefore to compactify the phase space, variable \(z\) is transformed as:
\begin{equation}
Z = \tan^{-1}z\,,
\end{equation}
where \(Z \) ranges from \(0 \le Z < \pi/2\). {Note that in the phase plot, we have only plotted some specific range of \(Z\) such that the \(Z>0\) plane must be visible in order to differentiate between the various trajectories.} In the given phase space some trajectories originate near \(P_{2}\) and initially get attracted towards \(P_{1}\). Near $P_1$ we have an effective matter dominated phase and the trajectories get repelled from it. After the repulsion the trajectories are attracted towards  \(P_{6}\) via \(P_{7}\).  	
The numerical evolution of cosmological variables is traced in Fig.[\ref{fig:evo_nonmin_quint_bet03}]. In the early phase both field energy density, total EoS, $\omega_{\rm tot}$, and field EoS, $\omega_{\phi}$, are near to one. As the interaction parameter \(z\) becomes non-negligible at \(N=0\), while matter density starts decreasing, the field EoS becomes \(-1\). In the late phase the field density dominates, $\omega_{\rm tot} \approx -1$, and fluid density becomes negligible. In the very early phase the interaction term \(z\) was negligible and $\sigma^2 \approx 0$. The interaction term becomes non-negligible during the late-time phase. This shows that in the NMC field-fluid system, during the late-time phase, the system never decouples and transfer of energy continuously takes place between these two dark sectors.\\

\begin{figure}[t]
	\begin{minipage}{0.5\linewidth}
		\centering
		\includegraphics[scale=0.74]{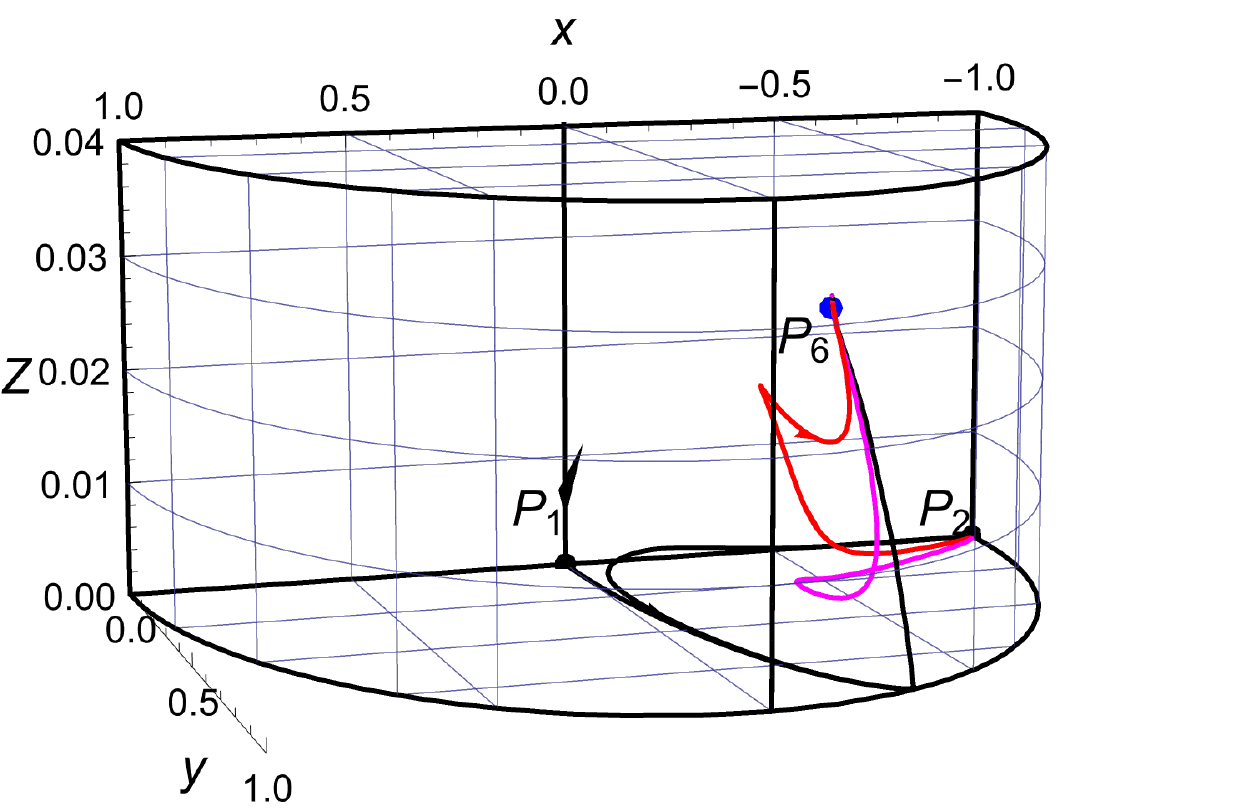}
		\caption{Phase space for $(\beta = 1, \gamma = -6.0, \lambda = 0.8, \alpha = 2.0)$. }
		\label{fig:phase_beta1_nonmin}
	\end{minipage}
	\hspace{0.2cm}
	\begin{minipage}{0.5\linewidth}
		\centering
		\includegraphics[scale=0.6]{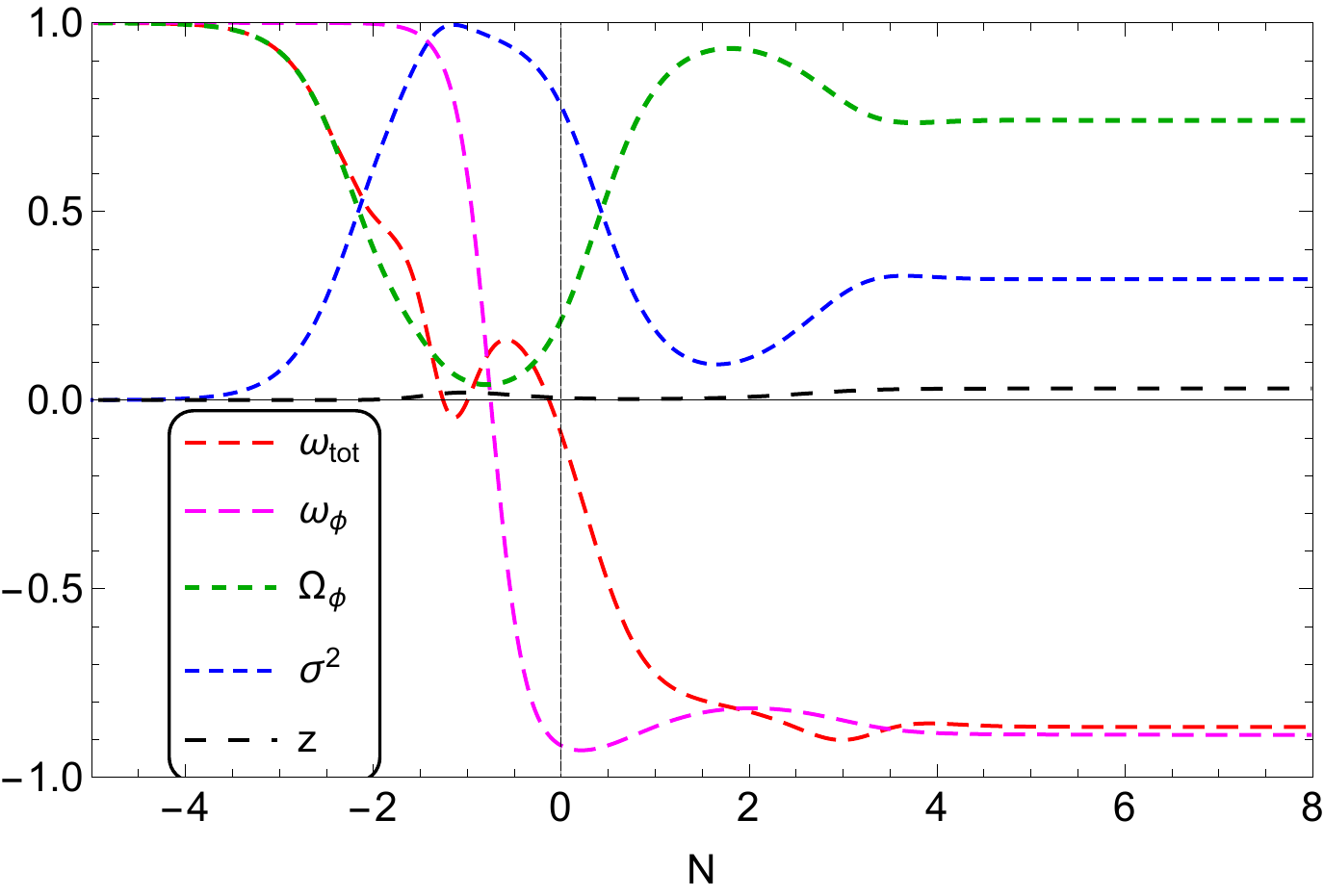}
		\caption{Numerical evolution for $(\beta = 1, \gamma = -6.0, \lambda = 0.8, \alpha = 2.0)$.}
		\label{fig:evo_beta1_nonmin}
	\end{minipage}
\end{figure}

\noindent{
\textbf{Case II: $(\beta = 1, \gamma = -6.0, \lambda = 0.8, \alpha = 2.0)$} In contrast to the previous case, the model yields only three physically viable critical points; we discover no additional critical points. \(P_{6}\) demonstrates the stable accelerating solution for this model, where field energy density dominates fluid energy density. At this point $\Omega_{\phi} \sim 0.75$, \(\sigma^2 \sim 0.32\) and $\omega_{\rm tot} \sim -0.90$. Thus, the interacting model \(f \propto \rho \) can generate an accelerating solution with non-zero fluid density $\simeq 0.30$ during the late-time phase. Note that in these two cases, one can transform the variable \((z \to 1/u)\) to obtain the critical points at infinity; however, from the Friedmann equation Eq.~\eqref{sigma_quint_non}, it becomes clear that any such transform that maps the \((z \to \infty, u \to 0) \), yields the matter-dominated solution irrespective of the accelerating or non-accelerating $(\omega_{\rm tot})$ characteristics. These critical points can not therefore be considered physically viable fixed points at the epoch of late-time cosmology.} 
The phase space and numerical evolution have been plotted in Fig.[\ref{fig:phase_beta1_nonmin}, \ref{fig:evo_beta1_nonmin}]. In the phase space some trajectories originate from \(P_{2}\) and some originate from some point just outside our region of interest. Initially these trajectories are attracted towards the matter dominated phase specified by the point \(P_{1}\) and then they get attracted towards \(P_{6}\). Thus $P_6$  becomes a global attractor in the phase space. Although both the models exhibit stable accelerating solution in the late-time phase yet the numerical evolution shows significant deviation in the latter case, see Fig.[\ref{fig:evo_beta1_nonmin}]. The evolution of $\omega_{\rm tot}$ shows some oscillatory behavior during the matter dominated phase, which shows that although the dark matter EoS is zero, yet, the coupled system can produce non-zero pressure. As the system enters the late time phase, the field density starts increasing and saturates to \(\approx 70\%\), while the fluid density becomes \(\approx 30\%\). In the late-time the $\omega_{\rm tot} $ approaches \(\approx - 1\).

In summary, the NMC field-fluid system can produce a stable accelerating expansion phase with total EoS \(-1\). All the results shown here point to a stiff matter dominated phase in the early phase, but one has the liberty to choose the initial phase. All the results show that the total EoS comes down as time evolves and one may choose the early phase (of the late universe) appropriately depending on the total EoS of the system. Additionally, we discover that for some interaction parameters, $\omega_{\rm tot}$ can cross $-1$, however we have not found any stable critical point showing phantom induced expansion. We can not rule out the possibility of determining the stable critical points for phantom case as for some other choices of parameters the complex system may give rise to such interesting critical points. 
	
\section{Conclusion}
\label{sec_conclusion}

Non-minimal coupling of scalar fields and fluids has been studied thoroughly in the context of late-time cosmology. These models provide an interacting dark sector. As presently our understanding of the basic constituents of the dark sector is not adequate, we do not know about any principles or rules which can forbid these non-minimal interactions. The most general action consisting of DM and DE admit non-minimal coupling term between them and consequently the NMC models are important models for the dark sector as long as they are not convincingly refuted by any observational evidence. Previously authors have also studied the non-minimal coupling of the DM sector with curvature in the cosmological context. Those models do not include DE and consequently cannot be taken as proper models which can address the cosmological dynamics of the late-time universe. In this paper, we have tried to address this issue.

Initially we have presented some models where the DM sector is non-minimally coupled to scalar curvature and the minimally coupled DE sector is produced by the quintessence like scalar field. In these models the DM and DE sectors do not directly couple to each other. Our primary aim in this study was to see whether we get stable critical points with accelerated expansion around them. Moreover, we preferred to have critical points where the non-minimal curvature coupling does not vanish. This is because a stable critical point around which the curvature coupling vanishes produces an accelerated expansion phase where the dark sector is uncoupled. Studies on uncoupled dark sectors have separately been done and consequently those results do not yield something very interesting. It is seen that the simplest model of DM and curvature coupling always produces critical points around which the non-minimal coupling vanishes. Although the dynamics of the universe in early phase is modified, in such a model, due to the presence of the non-minimal coupling the late time universe stable phase becomes uncoupled. This observation does not depend upon the nature of the scalar field, the decoupled nature of the critical points is observed for both quintessence or phantom like scalar fields.  It was observed that this particular nature of the critical points depended heavily on the form of the non-minimal coupling. In a different model, where the non-minimal coupling term was modified we obtained various critical points around which the curvature coupling of the DM sector never vanishes. In these cases, the energy density of the scalar field is modified due to the curvature coupling near the stable critical point. In the absence of the quintessence like scalar field, curvature coupling of the DM sector never produces any accelerated expansion solution. Only in the presence of the quintessence like scalar field one can get non-minimally interacting DM sector in the presence of accelerated expansion. Working with a quintessence like scalar field it was seen that the effective EoS never crossed the phantom line.   

The previous  model study produced an interesting question. If instead a curvature coupling of DM what will happen if we have the whole dark sector to be simultaneously coupled to curvature? In such a case the dark sector starts to interact with each other and this interaction is mediated by scalar curvature. We introduced such a type of interaction and found out the basic equations governing cosmological dynamics by choosing a particular form of the curvature coupling. The choice of the coupling term was made in such a manner so that the dynamical system remains relatively manageable. In reality, the simplest models of curvature coupling of the dark sector are intricate and we do not claim that we have exhaustively studied the system. We have simplified our model analysis by constraining the 3-dimensional phase space and we have also studied some particular kinds of critical points around which the coupling term does not vanish. Even in the simplest model the results are interesting. We show that we can obtain relevant stable fixed points around which we get accelerated expansion. Moreover, we show that even when one works with quintessence fields one can cross the phantom divide. This property is obtained because of the non-minimal interaction term.  

In conclusion we state that we have studied models of non-minimal interaction of the dark sector mediated by scalar curvature. All the basic results are obtained from an action principle and consequently the results are as general as they can be. Later on to produce cosmologically relevant results we have chosen various particular forms of non-minimal interaction. Throughout we have worked with the standard form of quintessence field potential. The results produce interesting late time cosmologies, while in the early phase most of the results predict a stiff matter dominated phase which slowly comes down. One can always modify the initial point and choose the particular effective EoS to work with. These models can have interesting observational signatures as the non-minimal coupling term will always modify the theory of structure formation. 
\vskip .5cm
\noindent {\bf Acknowledgement:\,\,} Authors are thankful to the referee for the valuable suggestions. A.C. would like to thank Indian Institute of Technology, Kanpur, for supporting this work by means of Institute Post-Doctoral Fellowship (Ref. No. DF/PDF197/2020-IITK/970). 

\appendix


\section{Variation of the fluid variables} 
\label{appen:1}

The variation of the action in Eq.\eqref{betoni_action} with respect to the fluid variables yields: 
\begin{eqnarray}
J^{\mu}  \ : \quad && \left(\frac{\p \rho }{\p n}  - \alpha \frac{\p f}{\p n } 
		\frac{R}{2 \kappa^2 } \right) U_{\mu} + \left(\varphi_{, \mu} + s \theta_{,\mu} + \beta_A \alpha^{A}_{, \mu}\right)=0, \\
		s \ : \quad && - \frac{\p \rho }{\p s} + \alpha \frac{\p f}{\p s}  \frac{R}{2 \kappa^2 } + n U^{\mu} \nb_{\mu }\theta = 0, \\
		\varphi \ : \quad && \nb_{\mu} J^{\mu } = 0, \\
		\theta \ : \quad && \nb_{\mu} (s J^{\mu }) = 0,\\
		\alpha^A \ : \quad && \nb_{\mu }(J^{\mu} \beta_A) = 0,\\
		\beta_A\ : \quad && J^{\mu } \nb_{\mu } \alpha^{A} = 0\,.
\end{eqnarray}
Here the interaction term \( f( n, s)\) in general only depends on the fluid parameters. However, if the interaction term also has dependence on other variables apart from the fluid parameters, as given in action in Eq.\eqref{action_coupled_quintessence} where \( f(n,s) \mapsto f(n,s,\phi)\), the above equations of the motion remains the same. Because of the structure of the relativistic fluid, the variation in $\varphi$ and $\theta$ puts additional constraints. The number density in the FLRW metric is conserved from \(\nb_{\mu} J^{\mu } = 0 \implies \dot{n} + 3 n H = 0 \). However, this constraint can be lifted by introducing a source term as discussed in \cite{Lima:2012cm,Bhattacharya:2020bwz}: 
	\begin{equation}\label{}
		\dot{n} + 3 n H = n \Gamma\,. 
	\end{equation}
	where \( \Gamma > 0 \) is a particle creation rate in a comoving volume \( a^3\). With such modification, the system becomes thermodynamically open and induces a negative creation pressure. Moreover, the system is thermodynamically adiabatic \( \nb_{\mu} (s J^{\mu }) = 0 \implies \dot{s} = 0 \), which means the constant entropy per particle. Therefore, one can investigate the non-minimal coupling of the fluid as an open thermodynamics system. The modified fluid equations can be identified as thermodynamic quantities such as temperature and chemical free energy \cite{Bettoni:2015wla,Boehmer:2015kta}.

{
\section{Critical points at infinity } 

\subsection{Non-minimally coupled fluid curvature system \label{appen:cric_infinity}}
To obtain the critical points at infinity, a simple transformation of the unconstrained variable \(z\) can be used to map the critical points at infinity to a finite value. The simplest transformation is 
\begin{equation}
	z \to 1/u\,.
	\label{map}
\end{equation}
This mapping allows us to shift the critical points from \(z = \infty \) to \(u=0\). With this transformation, the autonomous equations Eqs.~(\ref{z_prime_cf1}, \ref{z_prime_cf2}) for pressureless fluid becomes:
\begin{eqnarray}
	\text{Model I} \rightarrow \ 	u' &= & 3 \beta  u, \quad \text{Model II} \rightarrow  \ u' = 3 \beta u \ln|1/u|\,.
	\label{fluid_conf_trans}
\end{eqnarray}
Both of these autonomous equations have $u=0$ as a critical point. The fluid density and effective EoS for 
Model I are:
\begin{equation}
	\begin{split}
		\sigma^2  = 1-\alpha \frac{1}{u} (2+3(\beta-1)), \quad \omega_{\rm tot} = \frac{-7 \alpha  \beta }{-3 \alpha  \beta +2 \alpha +2 u}, 
	\end{split}
\end{equation}
and for Model II:
\begin{equation}
	\sigma^2  = 1-  \frac{3 \alpha \beta}{u}  \ln|1/u| + \alpha /{u}, \quad \omega_{\rm tot} = \frac{-7 \alpha  \beta  \log \left(\frac{1}{u}\right)}{2 (\alpha +u)-3 \alpha  \beta  \log \left(\frac{1}{u}\right)} \,.
\end{equation}
At \(u = 0\), the autonomous equations for both the models are regular, whereas the fluid density becomes singular.
This shows that fluid density dominates at this fixed point and can produce finite total equation of state. Model I can render the accelerated expansion solution \(-1< \omega_{\rm tot} <-1/3\) for \((\frac{1}{12}<\beta <\frac{1}{5})\) and phantom like solution\(-1.5<\omega_{\rm tot}<-1\) for \((\frac{1}{5}<\beta <\frac{6}{23})\). The point also features non-accelerating solution \(0 \le \omega_{\rm tot}\le 1/3\) for $-\frac{1}{9}\leq \beta \leq 0$. Therefore, a positive $\beta$ can produce an accelerating and phantom solution, but as \((N \to +\infty)\) the point becomes unstable. In contrast negative $\beta$ exhibits the non-accelerating attractor solution for \((N  \to +\infty)\). Model II, yields \(\lim\limits_{u \to 0} \omega_{\rm tot} = 7/3\), which produces non-accelerating stiff matter solution and the point turns out to be unstable.  The other critical point \(u_* = 1\) in model II, is equivalent to the \(z = 1\) point. This demonstrates that the model possesses no stable accelerating solutions at infinity. 
\subsection{Minimally coupled quintessence field \label{power_law_infinity}}

In all the models which include the quintessence scalar and matter, we see from the respective Friedmann constraints that for $\epsilon=-1$, $x$ and $y$ are bounded variables as if they tend to infinity $\sigma^2$ become negative. On the other hand for suitable choices of parameters one can take $z\to \infty$, keeping $\sigma^2>0$. This means only $z$ is unconstrained and reaches infinity. We will see that in almost all the cases as $z\to \infty$ we have $\sigma^2\to \infty$ which violates the bound $0\le \sigma^2 \le 1$. Although the above bound is violated we still present the nature of the fixed points at infinity for the sake of mathematical completeness. 

Using Eq.~\eqref{map} we can similarly extract the critical points corresponding to the interaction chosen in Eq.~\eqref{power_law}.
As a result of this redefinition, the system of autonomous equations Eq.~\eqref{eq:x_prime1}-\eqref{eq:z_prime1} for matter background  are modified as:
\begin{subequations}
	\begin{eqnarray}
		x' &=& \frac{2 u \left(3 x^3-3 x \left(y^2+1\right)+\sqrt{6} \lambda  y^2\right)-\alpha  \left(6 (2 \beta +1) x+\sqrt{6} (3 \beta -2) \lambda  y^2\right)}{\alpha  (4-6 \beta )+4 u} \label{trans_x_mini}\\
		y' &= & \frac{y \left(u \left(6 x^2-2 \sqrt{6} \lambda  x-6 y^2+6\right)+\alpha  \left(3 \beta  \left(\sqrt{6} \lambda  x-10\right)-2 \sqrt{6} \lambda  x+6\right)\right)}{\alpha  (4-6 \beta )+4 u} \label{trans_y_mini}\\
		u' &= & 3 \beta  u\,,
	\end{eqnarray}
\end{subequations}
and the corresponding effective EoS and fluid density becomes:
\begin{equation}
	\omega_{\rm tot} = \frac{-7 \alpha  \beta +2 u x^2-2 u y^2}{-3 \alpha  \beta +2 \alpha +2 u}, \quad \sigma^2 = -\frac{\alpha  (3 (\beta -1)+2)}{u}-x^2-y^2+1\,.XS
	\label{new_sigma}
\end{equation}
It turns out that \(u=0\) is a valid critical point for the 3D autonomous system. However, the fluid density \((\sigma^2) \) from Eq.~\eqref{new_sigma}, at \(u=0\) becomes singular but the effective equation of state $(\omega_{\rm tot})$ at \(u = 0\) becomes independent of \((x,y)\). The fixed points at infinity are tabulated in Tab.[\ref{tab:critical_pts_power_quint_infint}].
\begin{table}[t]
	{
	\small
	\centering
	\begin{tabular}{|c|c|c|c|c|c|c|c|}
		\hline
		Points & $ x $ & $ y $ & $ u $ & $\Omega_{\phi}$& $ \sigma^2 $ & $\omega_{\rm tot}$& Stability\\
		\hline 
		\hline
		$ P_{1} $ & 0 & 0 & 0 & 0 & $\infty$& $\frac{7 \beta }{3 \beta -2}$ & $ \makecell{\bigg( 3 \beta ,-\frac{3 \left(\sqrt{\alpha ^4 (2-3 \beta )^4}+7 \alpha ^2 \beta  (2-3 \beta )\right)}{2 \alpha ^2 (2-3 \beta )^2},\\ \frac{3 \left(\sqrt{\alpha ^4 (2-3 \beta )^4}+7 \alpha ^2 \beta  (3 \beta -2)\right)}{2 \alpha ^2 (2-3 \beta )^2} \bigg) }$ \\
		\hline 
		$ P_{2} $ & $\frac{\sqrt{6} (5 \beta -1)}{(3 \beta -2) \lambda }$ & $\frac{\sqrt{-60 \beta ^2-18 \beta +6}}{\sqrt{(2-3 \beta )^2 \lambda ^2}}$ & 0 & $\frac{6 (5 \beta -1)}{(3 \beta -2) \lambda ^2}$ & $\infty$ & $\frac{7 \beta }{3 \beta -2}$&  $\makecell{\bigg(  3 \beta ,-\frac{3 \left(\sqrt{7} \sqrt{\alpha ^4 (2-3 \beta )^2 (2 \beta +1) (6 \beta -1)}+\alpha ^2 \left(-6 \beta ^2+\beta +2\right)\right)}{2 \alpha ^2 (2-3 \beta )^2}, \\ \frac{3 \left(\sqrt{7} \sqrt{\alpha ^4 (2-3 \beta )^2 (2 \beta +1) (6 \beta -1)}+\alpha ^2 (2 \beta +1) (3 \beta -2)\right)}{2 \alpha ^2 (2-3 \beta )^2} \bigg) } $\\
		\hline 
		
	\end{tabular}
}
	\caption{Critical Points at infinity for minimally coupled quintessence field \((\epsilon = -1)\) with non-minimally coupled fluid-curvature interaction \((f = M^{-4\beta} \rho^{\beta} (n,s))\) for pressureless  background fluid.} 
	\label{tab:critical_pts_power_quint_infint}
\end{table}
The system yields two critical points and both of them have same total EoS which shows accelerated expansion phase \(-1< \omega_{\rm tot} <-1/3\) for \((\frac{1}{12}<\beta <\frac{1}{5})\), phantom phase \(-1.5<\omega_{\rm tot}<-1\) for \((\frac{1}{5}<\beta <\frac{6}{23})\) and non-accelerating expansion phase \(0 \le \omega_{\rm tot}\le 1/3\) for $-\frac{1}{9}\leq \beta \leq 0$. On finding the stability, one of the eigenvalues becomes positive (negative) for $\pm \beta$ and thus both the points become saddle. This explains the behavior of $\sigma^2$ and \(z\) at the past epoch i.e., negative \(N\) in Fig.[\ref{fig:evo_quint}]. Note that similar analysis can  also be carried out for the exponential interaction  case, as specified by Eq.~\eqref{int_expo_quint_mini}, and the corresponding autonomous equation Eq.~\eqref{exp_z_prime} becomes: 
\begin{equation}
	u' = 3 \beta u (1+ \omega) \ln|1/u|\,.
\end{equation}
The rest of the autonomous equations in \((x,y)\) remain same, as before, and given by Eqs.~(\ref{trans_x_mini}, \ref{trans_y_mini}).
For the transformed system, we get only one valid critical point for \((x=y=u=0)\). Hence, this becomes similar to the above case discussed in Eq.~\eqref{fluid_conf_trans}. Therefore, these analysis shows that the models does not produce any stable accelerating expansion phase.}


\end{document}